\documentclass[useAMS,usenatbib]{mn2e}

\usepackage{lastpage}
\usepackage{times,pslatex}
\usepackage{amsmath,amsfonts,graphicx}
\usepackage{color}

\newcommand{\Mjup}{\mbox{M$_{Jup}$}}

\newcommand{\Msun}{\mbox{M$_{\odot}$}}

\title[A Dynamical Analysis of the Proposed Circumbinary HW Virginis 
Planetary System]{A Dynamical Analysis of the Proposed Circumbinary HW 
Virginis Planetary System}

\author[J. Horner, T. C. Hinse, R.A. Wittenmyer, J. P. Marshall \& C.G. Tinney]{J. Horner$^{1}$\thanks{E-mail: j.a.horner@unsw.edu.au (JH)}, T. C. Hinse$^{2,3}$, R. A. Wittenmyer$^{1}$,  J. P. Marshall$^{4}$ \& C. G. Tinney$^{1}$ \\ \\
$^{1}$Department of Astrophysics and Optics, School of 
Physics, University of New South Wales, Sydney 2052, Australia\\ 
$^{2}$Korea Astronomy and Space Science Institute, 776 Daedeokdae-ro, Yuseong-gu, 305-348, Daejeon, Republic of Korea (South) \\
$^{3}$Armagh Observatory, College Hill, BT61 9DG, NI, UK \\
$^{4}$Departmento F\'isica Te\'orica, Facultad de Ciencias, Universidad 
Aut\'onoma de Madrid, Cantoblanco, 28049, Madrid, Espa\~na}
\begin{document}

\date{\Large{\color{red}{Accepted for publication in Monthly Notices of the Royal Astronomical Society, 3rd September 2012}}}

\pagerange{\pageref{firstpage}--\pageref{lastpage}} \pubyear{2012}

\maketitle
\vspace{-0.9cm}
\begin{abstract}

In 2009, the discovery of two planets orbiting the evolved binary star 
system HW Virginis was announced, based on systematic variations in the 
timing of eclipses between the two stars.  The planets invoked in that 
work were significantly more massive than Jupiter, and moved on orbits 
that were mutually crossing - an architecture which suggests that mutual 
encounters and strong gravitational interactions are almost guaranteed.

In this work, we perform a highly detailed analysis of the proposed 
HW~Vir planetary system.  First, we consider the dynamical stability of 
the system as proposed in the discovery work. Through a mapping process involving 91,125 individual simulations, we find that the system is so unstable that the planets proposed simply cannot exist, due to mean lifetimes of less 
than a thousand years across the whole parameter space.

We then present a detailed re-analysis of the observational data on 
HW~Vir, deriving a new orbital solution that provides a very good fit to 
the observational data. Our new analysis yields a system with planets more
widely spaced, and of lower mass, than that proposed in the discovery work,
and yields a significantly greater (and more realistic) estimate of the uncertainty in the orbit of the outermost body. Despite this, a detailed dynamical analysis of this new solution similarly reveals that it also requires the planets to move on orbits that are simply not dynamically feasible. 

Our results imply that some mechanism other than the influence of 
planetary companions must be the principal cause of the observed eclipse 
timing variations for HW~Vir. If the system does host exoplanets, they 
must move on orbits differing greatly from those previously proposed.  
Our results illustrate the critical importance of performing dynamical 
analyses as a part of the discovery process for multiple-planet 
exoplanetary systems.

\end{abstract}

\begin{keywords}
binaries: close, binaries: eclipsing, stars: individual: HW~Vir, planetary systems, 
methods: N-body simulations
\end{keywords}

\vspace{-1.0cm}

\section{Introduction}
\label{firstpage}

Since the discovery of the first planets around other stars 
\citep{WolszczanFrail1992, Mayor95}, the search for exoplanets has 
blossomed to become one of the most exciting fields of modern 
astronomical research.  The great majority of the hundreds of exoplanets 
that have been discovered over the past two decades have been found 
orbiting Sun-like stars by dedicated international radial velocity 
programs.  Among these programmes are HARPS \citep[the High Accuracy 
Radial velocity Planetary Search project, e.g.  ][]{Pepe04, Udry07, 
Mayor09}, AAPS \citep[the Anglo-Australian Planet Search, e.g.  
][]{Tinney01, Tinney11, Wittenmyer12b}, California \citep[e.g. 
][]{howard10, wright11}, Lick-Carnegie \citep[e.g. ][]{rivera10, ang12}, 
and Texas \citep[e.g. ][]{endl06, Robertson12a}.  The other main method 
for exoplanet detection is the transit technique, which searches for the 
small dips in the brightness of stars that result from the transit of 
planets across them. Ground-based surveys such as WASP \citep{WASP43, 
WASP36} and HAT \citep{HAT1, HAT17} have pioneered such observations, 
and resulted in the discovery of a number of interesting planetary 
systems. In the coming years, such surveys using space-based 
observatories will revolutionise the search for exoplanets. Indeed, a 
rapidly growing contribution to the catalogue of known exoplanets comes 
from the \textit{Kepler} spacecraft \citep[e.g.  
][]{Borucki11,Welsh12,kepler16}, which will likely result in the number 
of known exoplanets growing by an order of magnitude in the coming 
years.

In recent years, a number of new exoplanet discoveries have been 
announced featuring host stars that differ greatly from the Sun-like 
archetype that make up the bulk of detections.  The most striking of 
these are the circumbinary planets, detected around eclipsing binary 
stars via the periodic variations in the timing of observed stellar 
eclipses.  A number of these unusual systems feature cataclysmic 
variable stars, interacting binary stars composed of a white dwarf 
primary and a Roche lobe filling M star secondary.  Sharply-defined 
eclipses of the bright accretion spot, with periods of hours, can be 
timed with a precision of a few seconds.  In these systems, the eclipse 
timings are fitted with a linear ephemeris, and the residuals ($O-C$) 
are found to display further, higher-order variations.  These variations 
can be attributed to the gravitational effects of distant orbiting 
bodies which tug on the eclipsing binary stars, causing the eclipses to 
appear slightly early or late.  This light-travel time (LTT) effect can 
then be measured and used to infer the presence of planetary-mass 
companions around these highly unusual stars.  Some examples of 
circumbinary companions discovered in this manner include UZ~For \citep{Potter2011}, NN~Ser \citep{Beuermann2010}, DP~Leo \citep{qian10}, 
HU~Aqr \citep{Schwarz2009, Qian2011} and SZ~Her \citep{Hinse2012a,Lee2012}

The first circumbinary planets to be detected around hosts other than 
pulsars were those in the HW~Vir system \citep{Lee2009}, which features 
a subdwarf primary, of spectral class B, and a red dwarf companion which 
display mutual eclipses with a period of around 2.8~hours 
\citep{Menzies86}.  The detection of planets in this system was based on 
the timing of mutual eclipses between the central stars varying in a 
fashion that was best fit by including two sinusoidal timing variations.  
The first, attributed to a companion of mass $M~\sin~i\!=\!19.2~$\Mjup, 
had a period of 15.8~years, and a semi-amplitude of 77~s, while the 
second, attributed to a companion of mass $M~\sin~i\!=\!8.5~$\Mjup, had 
a period of 9.1~years and semi-amplitude of 23~s.  Whilst these 
semi-amplitudes might appear relatively small, the precision with which 
the timing of mutual eclipses between the components of the HW~Vir 
binary can be measured means that such variations are relatively easy to 
detect.

Over the last decade, a number of studies have shown that, for systems that are found to contain more than one planetary body, a detailed dynamical study is 
an important component of the planet discovery process that should not be 
overlooked \citep[e.g.][]{SMB00, GZ01, FMB05, LC09}. However, despite these pioneering works, the great majority of exoplanet discovery papers still fail to take account of the dynamical behaviour of the proposed systems. Fortunately, this situation is slowly changing, and recent discovery papers such 
as  \citet{Robertson12a}, \citet{Wittenmyer12b} and \citet{Robertson12b} have 
shown how studying the dynamical interaction between the 
proposed planets can provide significant additional constraints on the 
plausible orbits allowed for those planets. Such studies can even 
reveal systems in which the observed signal cannot be explained by the 
presence of planetary companions.  The planets proposed to orbit HU~Aqr 
are one such case, with a number of studies \citep[e.g.][]{Horner2011, FunkEPSC, 
Hinse2012, Wittenmyer12a, GZ12} showing that the orbital 
architectures allowed by the observations are dynamically unstable on 
astronomically short timescales.  In other words, whilst it is clear 
that the observed signal is truly there, it seems highly unlikely that it is solely the result of orbiting planets.

Given these recent studies, it is clearly important to consider whether 
known multiple-planet exoplanetary systems are truly what they seem to 
be.  In this work, we present a re-analysis of the 2-planet 
system proposed around the eclipsing binary HW~Vir. In section 2, we 
briefly review the HW~Vir planetary system as proposed by 
\citet{Lee2009}.  In section 3, a detailed dynamical analysis of the 
planets proposed in that work is performed.  In section 4, we present a 
re-analysis of the observations of HW~Vir that led to the 
announcements of the exoplanets, obtaining a new orbital solution for 
those planets which is dynamically tested in section 5. Finally, in 
section 6, we present a discussion of our work, and draw conclusions 
based on the results herein.
\vspace{-0.6cm}
\section{The HW~Vir planetary system}

The HW~Vir system consists of a subdwarf B primary and an M6-7 main 
sequence secondary.  The system eclipses with a period of 2.8~hours 
\citep{Menzies86}, and the stars have masses of $M_{1}=0.48$~\Msun\ and 
$M_{2}=0.14$~\Msun\ \citep{wood99}.  Changes in the orbital period of the 
eclipsing binary were first noted by \citet{kilkenny94}; further 
observations led other authors to suggest that the period changes were 
due to LTT effects arising from an orbiting substellar companion 
\citep{kilkenny03, iban04}. \citet{Lee2009} obtained a further 8~years 
of photometric observations of HW~Vir.  Those data, in combination with 
the previously published eclipse timings spanning 24~years, indicated 
that the period changes consisted of a quadratic trend plus two 
sinusoidal variations with periods of 15.8 and 9.1~years.  
\citet{Lee2009} examined alternative explanations for the cyclical 
changes, ruling out apsidal motion and magnetic period modulation via 
the Applegate mechanism \citep{app92}.  They concluded that the most 
plausible cause of the observed cyclic period changes is the 
light-travel time effect induced by two companions with masses 19.2 and 
8.5~\Mjup.  The parameters of their fit can be found in Table 
\ref{Lee09_params}.  Formally, the planets are referred to as 
HW~Vir~(AB)b and HW~Vir~(AB)c, but for clarity, we refer to the planets 
as HW~Vir~b and HW~Vir~c.

\begin{table}
\centering
\begin{tabular}{lccl}
\hline
Parameter & HW~Vir~b & HW~Vir~c & Unit \\
\hline
$M~\sin~i$ & 0.00809~$\pm$~0.00040 & 0.01836~$\pm$~0.000031 & $M_\odot$\\
$a~\sin~i$ & 3.62~$\pm$~0.52 & 5.30~$\pm$~0.23 & AU\\
$e$ & 0.31~$\pm$~0.15 & 0.46~$\pm$~0.05 & \\ 
$\omega$ & 60.6~$\pm$~7.1 & 90.8~$\pm$~2.8 & deg\\
$T$ & 2,449,840~$\pm$~63 & 2,454,500~$\pm$~39 & HJD\\
$P$ & 3316~$\pm$~80 & 5786~$\pm$~51 & days\\
%
%
\hline
\end{tabular}
\caption{Parameters for the two planetary bodies proposed in the HW~Vir system, taken from \citet{Lee2009} (their Table 7). }
\label{Lee09_params}
\end{table}

A first look at the fitted parameters for the two proposed planets 
reveals an alarming result: the planets are both massive, in the regime 
that borders gas giants and brown dwarfs, and occupy highly eccentric, 
mutually crossing orbits with separations that guarantee close 
encounters - generally a surefire recipe for dynamical instability (as 
seen for the proposed planetary system around HU~Aqr \citep[e.g. 
][]{Horner2011, Hinse2012, Wittenmyer12a, GZ12}.

\vspace{-0.6cm}
\section{A dynamical search for stable orbits}

The work by \citet{Lee2009} derived relatively high masses (8.5 and 19.2~\Mjup) 
and orbital eccentricities (0.31 and 0.46), and so significant mutual 
gravitational interactions are expected. To assess the dynamical stability of 
the proposed planets in the HW~Vir system, we 
performed a large number of simulations of the planetary system, 
following a successful strategy used on a number of previous studies 
\citep[e.g.][]{Marshall10, Horner2011, Wittenmyer12a, Wittenmyer12b, 
Robertson12a, Robertson12b, Horner12b}.  We used the Hybrid integrator 
within the n-body dynamics package MERCURY \citep{MERCURY-1, MERCURY-2}, 
and followed the evolution of the two giant planets proposed by 
\citet{Lee2009} for a period of 100 million years.  In order to examine 
the full range of allowed orbital solutions, we composed a 
grid of plausible architectures for the HW~Vir planetary system, each of 
which tested a unique combination of the system's orbital elements, 
spanning the $\pm$~3-$\sigma$ range in the observed orbital parameters.  
Following our earlier work, the initial orbit of HW~Vir~c (the 
planet with the best constrained orbit in \citet{Lee2009} was held fixed 
at its nominal best-fit values (i.e. $a = 5.3~$AU, $e = 0.46$ etc.).  
The initial orbit of HW~Vir~b was varied systematically such that 
the full 3-$\sigma$ error ellipse in semi-major axis, eccentricity, 
longitude of periastron and mean anomaly were sampled.  In our earlier 
work, we have found that the two main drivers of stability or 
instability were the orbital semi-major axis and eccentricity 
\citep[e.g.  ][]{Horner2011}, and so we sampled the 3-$\sigma$ region of 
these parameters in the most detail.

In total, 45 distinct values of initial semi-major axis were tested for 
the orbit of HW~Vir~b, equally distributed across the full 
$\pm~3$-$\sigma$ range of allowed values.  For each of these unique 
semi-major axes, 45 distinct eccentricities were tested, evenly 
distributed across the possible range of allowed values (i.e.  between 
eccentricities of 0.00 and 0.76).  For each of the 2025 $a-e$ pairs 
tested in this way, fifteen unique values of $\omega$ were tested (again 
evenly spread across the $\pm~3$-$\sigma$ range, whilst for each of the 
$a-e-\omega$ values, three unique values of mean anomaly were 
considered.  In total, therefore, we considered 91,125 unique orbital 
configurations for HW~Vir~b, spread in a $45 \times 45 \times 15 
\times 3$ grid in $a-e-\omega-M$ space.  In each of our simulations, the 
masses of the planets were set to their minimal $M sin i$ values, in 
order to maximise the potential stability of their orbits.  The orbital 
evolution of the planets was followed for a period of 100 million years, 
or until one of the planets was either ejected (defined by that planet 
reaching a barycentric distance of 20~AU), a collision between the 
planets occurred, or one of the planets collided with the central stars.  
If such a collision/ejection event occurred, the time at which it 
happened was recorded.

In this way, the lifetime of each of the unique systems was determined.  
This, in turn, allowed us to construct a map of the dynamical stability 
of the system, which can be seen in Fig~\ref{OriginalStability}.  As can 
be seen in that figure, none of the orbital solutions tested were 
dynamically stable, with few $a$-$e$ locations displaying mean lifetimes 
longer than 1,000 years. \footnote{The four yellow/orange/red hotspots 
in that plot between $a \sim 3.3$ and $a \sim 4.2$ AU are the result of 
four unusually stable runs, with lifetimes of 120~kyr ($a = 3.31~AU, e = 
0.12$), 250~kyr ($a = 3.74~AU, e = 0.22$), 56~kyr ($a = 4.02~AU, e = 
0.19$) and 49~kyr ($a = 4.24~AU, e = 0.05$). Such ``long-live'' outliers are 
not unexpected, given the chaotic nature of dynamical interactions, but 
given the typically very short lifetimes observed can significantly 
alter the mean lifetime in a given bin.}
  
Remarkably, we find that the proposed orbits for the HW~Vir 
planetary system are even less dynamically stable than those proposed 
for the now discredited planetary system around HU~Aqr 
\citep{Horner2011,Wittenmyer12a,Hinse2012,Horner12b, GZ12}.  Simply put, our 
result proves conclusively that, if there are planets in the HW~Vir 
system, they must move on orbits dramatically different to those proposed 
by \citet{Lee2009}. This instability is not particularly surprising, 
given the high orbital eccentricity of planet c, which essentially 
ensures that the two planets are on orbits that intersect one another, 
irrespective of the initial orbit of planet b.  Given that the two 
planets are not trapped within mutual mean-motion resonance (MMR), such an 
orbital architecture essentially guarantees that they will experience 
strong close encounters within a very short period of time, ensuring the 
system's instability.

\vspace{-0.3cm}
\section{Eclipse timing data analysis and LTT model}

Given the extreme instability exhibited by the planets proposed by 
\citet{Lee2009}, it seems reasonable to ask whether a re-analysis of the 
observational data will yield significantly different (and more 
reasonable) orbits for the planets in question.  We therefore chose to 
re-analyse the observational data, following a similar methodology as 
applied in an earlier study of HU~Aqr \citep{Hinse2012}.

At the basis of our analysis we use the combined mid-eclipse timing 
data set compiled by \citet{Lee2009}, including the times of secondary 
eclipses.  The timing data used in \cite{Lee2009} were recorded in 
the UTC (Coordinated Universal Time) time standard, which is known 
to be non-uniform \citep{Bastian2000,GuinanRibas2001}. To eliminate timing 
variations introduced by accelerated motion within the Solar System, 
we therefore transformed\footnote{http://astroutils.astronomy.ohio-state.edu/time} the HJD (Heliocentric Julian Date) timing 
records in UTC time standard into Barycentric Julian Dates (BJD) 
within the Barycentric Dynamical Time (TDB) standard 
\citep{Eastman2010}. A total of 258 timing measurements were used 
spanning 24~years from January 1984 (HJD 2 445 730.6) to May 2008 
(HJD 2 454 607.1). We assigned 1-$\sigma$ timing uncertainties to each 
data point by following the same approach as outlined in \cite{Lee2009}.

For an idealised, unperturbed and isolated binary system, the linear 
ephemeris of future/past mid-eclipse (usually primary) events can be 
computed from
\begin{equation}
T_{C}(E) = T_{0} + P_{0}E,
\label{ephemeriseq}
\end{equation} 
\noindent
where $E$ denotes the (independent) ephemeris cycle number, $T_{0}$ is 
the reference epoch, and $P_{0}$ measures the eclipsing period ($\simeq 
2.8$ hrs) of HW~Vir.  A linear regression performed on the 258 recorded 
light-curves allows $P_{0}$ to be determined with high precision.  In 
this work, we chose to place the reference epoch close to the middle of 
the observing baseline to avoid parameter correlation between $T_{0}$ 
and $P_{0}$ during the fitting process. In the following we briefly 
outline the LTT model as used in this work.

\vspace{-0.3cm}
\subsection{Analytic LTT model}

The model adopted in this work is similar to that described in 
\cite{Hinse2012}, and is based on the original formulation of a single 
light-travel time orbit introduced by \cite{Irwin1952}. In this model the two components of the binary system are assumed to represent one 
single object with a total mass equal to the sum of the masses of the 
two stars. This point mass is then placed at the original binary 
barycentre. If a circumbinary companion exist, then the combined binary 
mass follows an orbit around the total system barycentre. The eclipses are then 
given by Eq.~\ref{ephemeriseq}. This defines the LTT orbit of the binary. The underlying reference system has its origin at the total centre of mass.

Following \cite{Irwin1952}, if the observed mid-eclipse times exhibit a 
sinusoidal-like variation (due to one or more unseen companion(s)), then the 
quantity $O-C$ defines the light-travel time effect and is given by

\begin{equation}
(O-C)(E) = T_{O}(E) - T_{C}(E) = \sum_{i=1}^2\tau_{i},
\end{equation}
\noindent
where $T_{O}$ denotes the measured time of an observed mid-eclipse, and $T_C$ is the computed time of that mid-eclipse based on a linear ephemeris. We note that $\tau_1+\tau_2$ is the combined LTT effect from two \emph{separate} two-body LTT orbits.  The quantity $\tau_i$ is given by the following expression for each companion 
\citep{Irwin1952}:
\begin{equation}
\tau_{i} = K_{b,i}\Big[\frac{1-e_{b,i}^2}{1+e_{b,i}\cos f_{b,i}} 
\sin (f_{b,i}+\omega_{b,i}) + e_{b,i}\sin \omega_{b,i}\Big ],
\label{taueq}
\end{equation}
\noindent
where $K_{b,i} = a_{b,i}\sin I_{b,i}/c$ is the semi-amplitude of the 
light-time effect (in the $O-C$ diagram) with $c$ measuring the speed of 
light and $I_{b,i}$ is the line-of-sight inclination of the LTT orbit 
relative to the sky plane, $e_{b,i}$ the orbital eccentricity, $f_{b,i}$ 
the true longitude and $\omega_{b,i}$ the argument of pericenter of the 
LTT orbit.  The 5 model parameters for a single LTT orbit are given by 
the set $(a_{b,i}\sin I_{b,i}, e_{b,i}, \omega_{b,i}, T_{b,i}, 
P_{b,i})$.  The time of pericentre passage $T_{b,i}$ and orbital period 
$P_{b,i}$ are introduced through the expression of the true longitude as a 
time-like variable via the mean anomaly $M = n_{b,i}(T_{O}-T_{b,i})$, 
with $n_{b,i} = 2\pi/P_{b,i}$ denoting the mean motion of the combined 
binary in its LTT orbit.  Computing the true anomaly as a function of 
time (or cycle number) requires the solution of Kepler's equation.  We 
direct the interested reader to \cite{Hinse2012} for further details.

In Eq.~\ref{taueq}, the origin of the coordinate system is placed at the 
centre of the LTT orbit \citep[see e.g.][]{Irwin1952}.  A more natural 
choice (from a dynamical point of view) would be to use the system 
centre of mass as the origin of the coordinate system.  However, the 
derived Keplerian elements are identical in the two coordinate systems \citep[e.g.][]{Hinse2012a}. Finally, we note that our model does not include mutual gravitational interactions. We also only consider the combination of two 
LTT orbits from two circumbinary companions.

From first principles, some similarities exist between the LTT orbit and 
the orbit of the circumbinary companion.  First, the eccentricities 
($e_{b,i} = e_{i}$) and orbital periods ($P_{b,i}=P_{i}$) are the same.  
Second, the arguments of pericenter are $180^{\circ}$ apart from one 
another ($\omega_i = 180^{\circ} - \omega_{b,i}$). Third, the times of 
pericenter passage are also identical ($T_{b,i} = T_{i}$).

Information of the mass of the unseen companion can be obtained from the 
mass function given by
\begin{equation}
f(m_{i}) = \frac{4\pi^2(a_{b,i}\sin I_{b,i})^3}{G P_{b,i}^2} = 
\frac{4\pi^2(K_{b,i}c)^3}{G P_{b,i}^2} = 
\frac{(m_{i}\sin I_{i})^3}{(m_b + m_i)},~~~~~i = 1,2
\label{massfunction}
\end{equation}
\noindent
The least-squares fitting process provides a measure for $K_{b,i}$ and 
$P_{b,i}$, and hence the minimum mass of the companion can be found from 
numerical iteration.  In the non-inertial astrocentric reference frame, 
with the combined binary mass at rest, the companion's semi-major axis 
\emph{relative to the binary} is then calculated using Kepler's third 
law.

In \cite{Lee2009} the authors also accounted for additional period 
variations due to mass transfer and/or magnetic interactions between the 
two binary components.  These variations usually occur on longer time 
scales compared to orbital period variations due to unseen companions.  
Following \cite{Hilditch2001}, the corresponding ephemeris of calculated 
times of mid-eclipses then takes the form
\begin{equation}
T_{C} = T_{0} + P_{0} E + \beta E^2 + \tau_i, ~~~~~i = 1,2
\label{anothereq}
\end{equation}
\noindent
where $\beta$ is an additional free model parameter and accounts for a 
secular modulation of the mid-eclipse times resulting from interactions 
between the binary components.  Assuming the timing data 
of HW~Vir are best described by a two-companion system, and to be 
consistent with \cite{Lee2009}, we have used Eq.~\ref{anothereq} as our 
model which consists of 13 parameters.

\vspace{-0.5cm}
\section{Methodology and results from $\chi^2$-parameter search}

To find a stable orbital configuration of the two proposed circumbinary 
companions, we carried out an extensive search for a best-fit in $\chi^2$ 
parameter space. The analysis, methodology and technique follow the same 
approach as outlined in \cite{Hinse2012}. Here we briefly repeat the most 
important elements in our analysis.

We used the Levenberg-Marquardt least-square minimisation algorithm as 
implemented in the {\sc IDL}\footnote{The acronym {\sc idl} stands for 
Interactive Data Language and is a trademark of ITT Visual Information 
Solutions.  For further details see {\tt 
http://www.ittvis.com/ProductServices/IDL.aspx}.}-based software package 
{\sc MPFIT}\footnote{http://purl.com/net/mpfit} \citep{Markwardt2009}.  
The goodness-of-fit statistic $\chi^2$ of each fit was evaluated from the 
weighted sum of squared errors. In this work, we use the 
\emph{reduced} chi-square statistic $\chi^2_r$ which takes into account 
the number of data points and the number of freely varying model parameters.

We seeded 28,201 initial guesses within a Monte Carlo experiment. Each guess 
was allowed a maximum of 500 iterations before termination, with all 13 
model parameters (including the secular term) kept freely varying. Converged 
solutions with $\chi^2_r \le 10$ were accepted with the initial guess and 
final fitting parameters recorded to a file. After each converged iteration 
we also solved the mass function (Eq.~\ref{massfunction}) for the companion's 
minimum mass and calculated the semi-major axis relative to the system 
barycentre from Kepler's third law.

Initial guesses of the model parameters were chosen at random following 
either a uniform or normal distribution.  For example, initial orbital 
eccentricities were drawn from a uniform distribution within the 
interval $e \in [0.0, 0.8]$. Our initial guesses for the orbital 
periods were guided by a Lomb-Scargle (LS) discrete Fourier 
transformation analysis on the complete timing data set.  For the LS 
analysis we used the {\sc PERIOD04} software package 
\citep{LenzBreger2005} capable of analysing unevenly sampled data sets.  
Fig.~\ref{HWVirPowerSpec} shows the normalised LS power-spectrum.  The 
LS algorithm found two significant periods with frequencies 
$f_1 \simeq 1.4 \times 
10^{-4}~\textnormal{cycles/day}$ and $f_2 \simeq 2.1\times 
10^{-4}~\textnormal{cycles/day}$.  These frequencies correspond to 
periods of 7397 and 4672 days, respectively.  Hence the short-period 
variation is covered more than twice during the observing interval. Due 
to a lower amplitude, it contains less power within the data set.  Our 
random initial guesses for the companion periods were then drawn from a 
Gaussian distribution centred at these periods with standard deviation 
of $\pm~5$~years.  We call this approach a ``quasi-global'' search of 
the underlying $\chi^2$-parameter space.
\vspace{-0.3cm}
\subsection{Results - finding best fit and confidence levels}

Our best fit model resulted in a $\chi^2_r = 0.943$ and is shown in 
Fig.~\ref{HWVirBestFit} along with the LTT signal due to the inner and 
outer companion and the secular term. The corresponding root-mean-square 
(RMS) scatter of data around the best fit is 8.7 seconds, which is close 
to the RMS scatter reported in \cite{Lee2009}. The fitted model elements and derived quantities of our best fit are shown in Table~\ref{bestfitparamtable}. Compared to the system in \cite{Lee2009} we note that we now obtain a lower eccentricity for the inner companion and a larger eccentricity for the outer companion.  Furthermore, our two-companion system has also slightly expanded, with larger semi-major axes (and therefore longer orbital periods) for both companions compared to \cite{Lee2009}.

The next question to ask is how reliable or significant our best-fit 
solution is in a statistical sense. Assuming that the errors are normally 
distributed one can establish confidence levels for a multi-parameter fit 
\citep{Bevington1992}. We therefore carried out detailed two-dimensional 
parameter scans covering a large range around the best-fit value in 
order to study the $\chi^2_r$-space topology in more detail. In particular, we explored relevant model parameter combinations including $T_0, P_0$ and $\beta$.

In all our experiments we allowed the remaining model parameters to vary 
freely while fixing the two parameters of interest in the considered 
parameter range \citep{Bevington1992,Press1992}. Assuming parameter errors 
are normally distributed, our 1-, 2- and 3-$\sigma$ level curves provide 
the 68.3, 95.4 and 99.7\% confidence levels relative to our best-fit, 
respectively. In Fig.~\ref{3by2plot} we show a selection of our 
two-dimensional parameter scan considering various model parameters. 

Ideally, one would aim to work with parameters with little correlation 
between the two parameters. The lower-right panel in Fig.~\ref{3by2plot} 
shows the relationship between $T_0$ and $P_0$. The near circular shape of the level curves reveals that little correlation between the two parameters exists. This is most likely explained by our choice to locate the reference epoch in the middle of the dataset. The remaining panels in Fig.~\ref{3by2plot} show some correlations between the parameters. However, we have some indication of an unconstrained outer orbital period in the lower-left panel of Fig.~\ref{3by2plot}. We show the location of orbital mean motion resonances in the $(P_{1},P_{2})$ plane. Our best fit is located close to the 2:1 mean motion resonance.

To demonstrate that the outer period is unconstrained we have generated 
$\chi^2_r$-parameter scans in the $(P_{2},e_{2})$-plane as shown in 
Fig.~\ref{3by1plot}. Our best fit model is shown in the left-most panel 
of Fig.~\ref{3by1plot}, along with the 1-,2- and 3-$\sigma$ confidence 
levels. It is readily evident that the 1-$\sigma$ confidence level 
does not simply surround our best-fit model in a confined or ellipsoidal manner. 
We rather observe that all three level curves are significantly stretched 
towards solutions featuring longer orbital periods for the outer companion. This is demonstrated in the middle and right panels of Fig.~\ref{3by1plot}. Any longer orbital period for the outer companion therefore results in a $\chi^2_r$ with similar statistical significance as our best-fit model. 

In addition, we studied the $(a\sin i,P)$-parameter plane for the 
outer companion, as shown in the lower-middle panel of Fig.~\ref{3by2plot}. 
Here, we also observe that $a\sin i$ is unconstrained. The 
unconstrained nature of the outer companion's $a\sin i$ and orbital period has 
a dramatic effect on the derived minimum mass and the corresponding error 
bounds.
\vspace{-0.3cm}
\subsection{Results - parameter errors}

When applying the LM algorithm formal parameter errors are obtained 
from the best-fit covariance matrix. However, in our study, we sometimes 
encounter situations where some of the matrix elements become zero, or have singular values. However, at others times, the error matrix is returned with non-zero elements. In those cases we have observed that the outer orbital period is often better determined than that of the inner companion. In the case of 
HW~Vir, such solutions are clearly incongruous, given the relatively poor orbital characterisation of the outer body compared to that of the innermost.

Being suspicious about the formal covariance errors, we have resorted to 
two other methods to determine parameter errors. First, we attempted to 
determine errors by the use of the bootstrap method \citep{Press1992}. 
However, we found that the resulting error ranges are comparable to the 
formal errors extracted from the best-fit covariance matrix. Furthermore, 
the bootstrap error ranges were clearly incompatible with the 1-$\sigma$ error 
``ellipses'' discussed above. For example, from the top-left panel in Fig.~\ref{3by2plot} 
we estimate the 1-$\sigma$ error on the inner orbital period to be on the 
order of 50-100 days. In contrast, the errors for the inner orbital period obtained from our 
bootstrap method were of order just a few days. For this reason, we consider the bootstrap method 
to have failed, and it has therefore not been investigated 
further in this study. However, it is interesting to speculate on the possibility of dealing 
with a dataset which is characterised by ``clumps of data'', as seen in 
Fig.~\ref{HWVirBestFit}. When generating random (with replacement) bootstrap ensembles, there is the possibility that \emph{only a small variation} is being introduced for each random draw, as a result of the clumpiness of the 
underlying dataset. That clumpiness might be mitigated for, and the bootstrap method rendered still viable for the establishment of reliable errors, by enlarging the number of bootstrap ensembles to compensate for the lack of variation within single bootstrap data sets. One other possibility would be to replace clumps of data by a single data point reflecting the average of the clump. However, we have instead invoked a different approach.

Having located a best-fit minimum, we again seeded a large number of initial 
guesses around the best-fit parameters. This time, we considered only a 
relatively narrow range around the best-fit values (e.g. as shown in 
Fig.~\ref{3by2plot} and Fig.~\ref{3by1plot}. This ensured that the LM algorithm 
would iterate towards our best-fit model depending on the 
underlying $\chi^2_r$ topology and inter-parameter correlations. The initial 
parameter guesses were randomly drawn from a uniform distribution 
within a given parameter interval. We then iterated towards a best-fit value 
using LM, and recorded the best-fit parameters along with the corresponding 
$\chi^2_r$.

Generating a large number of guesses enabled us to establish statistics on the final best-fit parameters with $\chi^2_r$ within 1-, 2-, and 3-$\sigma$ confidence levels. We therefore performed a Monte Carlo experiment that considered several tens of thousands of guesses. To establish 1-$\sigma$ error bounds (assuming a normal distribution for each parameter), we then considered only those models that yielded $\chi^2_r$ within the 1-$\sigma$ confidence limits (inner level curves), as shown in Fig.~\ref{3by2plot} and Fig.~\ref{3by1plot}. The error for a given parameter is then obtained from the mean and standard deviation, and listed in Table~\ref{bestfitparamtable}. In order to test our assumption of normally distributed errors, we plotted histograms for the various model parameters in Fig.~\ref{ErrorHistogram}. For each histogram distribution, we fitted a Gaussian and established the corresponding mean and standard deviation. While some parameters follow a Gaussian distribution (for example the outer companion's eccentricity), other parameters show no clear sign of ``Gaussian tails''. This is especially true for the orbital period of the outer companion. Following two independent paths of analysis, we have demonstrated that the outer period is unconstrained, based on the present dataset, and should be regarded with some caution. However, we also point out a short-coming of our method of determining random parameter errors. The parameter estimates depend on the proximity of the starting parameters to the best-fit parameter. In principle, our method of error determination assumes that the best-fit model parameters are well-determined in terms of well-established closed-loop confidence levels around the best-fit parameters. The true random parameter error distribution for the two ill-constrained parameters might turn out differently. To put stronger constraints on the model parameters is clearly only possible by augmenting the existing timing data through a program of continuous monitoring of HW~Vir over the coming years \citep{Pribulla2012, Kona12}. As more data is gathered, the confidence levels in Fig.~\ref{3by1plot} will eventually narrow down.

\vspace{-0.5cm}
\section{The $\beta$ coefficient and angular momentum loss}

One of the features of our best-fit orbital solution is that it results in a relatively large $\beta$ coefficient, which can be related to a change in the period of the binary resulting from additional, non-planetary effects. Potential causes of such a period change include mass transfer, loss of angular momentum, magnetic interactions between the two binary components and/or perturbations from a third body on a distant and unconstrained orbit. In this study, the $\beta$ factor represents a constant binary period change \citep[see][p. 171]{Hilditch2001} with a linear rate of $dP/dt = -9.57 \times 10^{-9} ~\textnormal{days/yr}$, which is about 15 percent larger than the value reported in \citet{Lee2009}. We retained the $\beta$ coefficient in our model in order that our treatment be consistent with that detailed in \citet{Lee2009}, such that our results might be directly compared to their work. 

\citet{Lee2009} examined a number of combinations of models that incorporated a variety of potential causes for the observed period modulation. They found that the timing data is best described by two LTT and a quadratic term in the linear ephemeris model. In their work, \citet{Lee2009} carefully examined the contribution of period modulation by various astrophysical effects. They were able to provide arguments that rule out the operation of the Applegate mechanism, due to the lack of small-scale variations in the observed luminosities that would have an influence on the $J_2$ oblateness coefficient of the magnetically active component. A change in $J_2$ would, in turn, affect the binary period. 

Furthermore, \citet{Lee2009} reject the idea that the observed $O-C$ variation could be the result of apsidal motion, based on the circular orbit of the HW~Vir binary system. In addition, they estimated the secular period change of the HW~Vir binary orbit due to angular momentum loss through gravitational radiation and magnetic breaking. They found that the most likely explanation for the observed linear decrease in the binary period is that it is the result of angular momentum loss by magnetic stellar wind breaking in the secondary M-type component. From first principles \citep[e.g.][]{Brinkworth2006}, the period change observed in this work corresponds to a angular momentum change of order $dJ/dt = -2.65 \times 10^{36}$ erg. This is approximately 15 percent larger than the value reported by \citet{Lee2009}, but still well within the range where magnetic breaking is a reasonable astrophysical cause for period modulation. 

Finally, we note that, whilst this work was under review, \citet{Beuermann2012} published a similar study based on new timing data, in which they also considered the influence of period changes due to additional companions. In their work, they obtained a markedly different orbital solution than those discussed in this work, one which they found to be dynamically stable on relatively long timescales. In light of their findings, it is interesting to note that they do not include a quadratic term in their linear ephemeris model. This could point at the possibility that the inclusion of a quadratic term is somehow linked to the instability of the best-fit system found in this work.

\citet{Beuermann2012} found stable orbits for a solution involving two circumbinary companions. However, despite this, we note that the two models share some qualitative characteristics. A careful examination of Fig.~2 in \citet{Beuermann2012} reveals that the orbital period of the outer companion is unconstrained from a period analysis since the $\chi^2$-contour curves are 
open towards longer outer orbital periods - a result mirrored in our current work. As we noted earlier, we were unable to place strict confidence levels on the best-fit outer companion's orbital period and semi-major axis. Although \citet{Beuermann2012} do find a range of stable scenarios featuring their outer companion, we note that they fixed the eccentricity of that companion's orbit to be near-circular, with period of 55~years. Such an assumption (i.e. fixing some orbital parameters) is somewhat dangerous, since it can lead to the production of dynamically stable solutions that are not necessarily supported by the observational data \citep[e.g.][]{Horner12b}. A more rigorous strategy would be to generate an ensemble of models with each model (all parameters freely varying) tested for orbital stability (using some criterion like non-crossing orbits or non-overlap of mean-motion resonances, etc.) resulting in a distribution of stable best-fit models. Using the new data set, a study of the distribution of the best-fit outer planet's eccentricity would be interesting. It is certainly possible that the new data set constrains this parameter sufficiently in order to validate their assumptions. Although the results presented in \citet{Beuermann2012} are clearly promising, it is definitely the case that more observations are needed before the true origin of the observed variation for HW~Vir is established beyond doubt.

The question of whether a period damping factor is truly necessary for the HW~Vir system would require a statistically self-consistent re-examination of the complete data set taking account of a range of model scenarios. We refer the interested reader to \citet{GZ12}, who recently carried out a detailed investigation of the influence of the quadratic term for various scenarios in their attempt to explain the timing data of the HU Aqr system.

\vspace{-0.5cm}
\section{Dynamical analysis of the best-fit LTT model}

Since a detailed re-analysis of the observational data on the HW~Vir 
system yields a new orbital solution for the system, it is interesting 
to consider whether that new solution offers better prospects for 
dynamical stability than that proposed in \citet{Lee2009}.  We therefore 
repeated our earlier dynamical analysis using the new orbital solution.  
Once again, we held the initial orbit of planet HW~Vir~c fixed at 
the nominal best-fit solution, and ran an equivalent grid of unique 
dynamical simulations of the planetary system, varying the initial orbit 
of HW~Vir~b such that a total of 45 distinct values of $a$ and $e$, 
15 distinct values of $\omega$ and 3 values of $M$ were tested, each 
distributed evenly as before across the $\pm~3$-$\sigma$ range of 
allowed values.  As before, the two simulated planets were assigned the 
nominal $M~\sin~i$ masses obtained from the orbital model.  The results 
of our simulations can be seen in Fig.~\ref{DynamicsNew}.

As was the case for the original orbital solution proposed in 
\citet{Lee2009}, and despite the significantly reduced uncertainties in 
the orbital elements for the resulting planets, very few of the tested 
planetary systems survived for more than 1,000~years (with just twenty six 
systems, 0.029\% of the sample, surviving for more than 3,000~years, and 
just three systems surviving for more than 10,000~years).  As was the case 
for the planetary system proposed to orbit the cataclysmic variable 
system HU~Aqr \citep[e.g. ][]{Horner2011, Wittenmyer12a}, it seems almost 
certain that the proposed planets in the HW~Vir system simply do not 
exist - at least on orbits resembling those that can be derived from the 
observational data.

\vspace{-0.5cm}
\section{Conclusion and Discussion}

The presence of two planets orbiting the evolved binary star system 
HW~Vir was proposed by \citet{Lee2009}, on the basis of periodic 
variations in the timing of eclipses between the two stars.  The planets 
proposed in that work were required to move on relatively eccentric 
orbits in order to explain the observed eclipse timing variations, to 
such a degree that the orbit of the outer planet must cross that of the 
innermost.  It is obvious that, when one object moves on an orbit that 
crosses that of another, the two will eventually encounter one another, 
unless they are protected from such close encounters by the influence of 
a mutual mean-motion resonance \citep[e.g.  ][]{Horner04a, Horner04b}.  
Even objects protected by the influence of such resonances can be 
dynamically unstable, albeit on typically longer timescales \citep[e.g.  
][]{HL10, HL12b, HL12c}. Since the two planets proposed by \citet{Lee2009} move on calculated orbits that allow the them to experience close encounters and yet are definitely not protected from such encounters by the influence of mutual mean-motion resonance, it is clear that they are likely to be highly 
dynamically unstable. To test this hypothesis, we performed a suite of 
highly detailed dynamical simulations of the proposed planetary system 
to examine its dynamical stability as a function of the orbits of the 
proposed planets. We found the proposed system to be dynamically 
unstable on extremely short timescales, as was expected based on the 
proposed architecture for the system.

Following our earlier work \citep{Wittenmyer12a, Hinse2012}, we 
performed a highly detailed re-analysis of the observed data, in order 
to check whether such improved analysis would yield better constrained 
orbits that might offer better prospects for dynamical stability.  Our 
analysis resulted in calculated orbits for the candidate planets in the 
HW~Vir system that have relatively small uncertainties.  Once these 
orbits had been obtained, we performed a second suite of detailed 
dynamical simulations to ascertain the dynamical stability of the newly 
determined orbits.  Following the same procedure as for the original 
orbits, we considered the stability of all plausible architectures for 
the HW~Vir system.  Despite the increased precision of the newly 
determined orbits, we find that the planetary system proposed is 
dynamically unstable on timescales as short as a human lifetime. For 
that reason, we must conclude that the eclipse-timing variations observed 
in the HW~Vir system are not solely down to the gravitational influence of 
perturbing planets. Furthermore, if any planets do exist in that system, they must move on orbits dramatically different to those considered in this work.

Our results highlight the importance of performing complementary 
dynamical studies of any suspected multiple-exoplanet system -- 
particularly in those cases where the derived planetary orbits approach 
one another closely, are mutually crossing and/or derived companion 
masses are large. Following a similar strategy as applied to the 
proposed planetary system orbiting HU~Aqr \citep{Horner2011, 
Wittenmyer12a, GZ12}, we have found that the proposed 
2-planet system around HW~Vir does not stand up to a detailed dynamical 
scrutiny. In this work we have shown that the outer companion's 
period (among other parameters) is heavily unconstrained by establishing 
confidence limits around our best-fit model. However, we also point out 
the fact that the two circumbinary companions have brown-dwarf masses. 
Hence, a more detailed $n$-body LTT model which takes account of
mutual gravitational interactions might provide a better description 
of the problem. 

To further characterise the HW~Vir system and constrain orbital 
parameters we recommend further observations within a monitoring 
program as described in \citet{Pribulla2012}. In a recent work on HU Aqr, 
\citet{GZ12} pointed out the possibility that different data sets obtained 
from different telescopes could introduce systematic errors resulting in 
a false-positive detection of a two-planet circumbinary system.

Finally, we note that, whilst this paper was under referee, \citet{Beuermann2012} independently published their own new study of the HW~Vir system. Based on new observational timing data, those authors determined a new LTT model that appears to place the 2-planet system around HW~Vir on orbits that display long-term dynamical stability. Based on the results presented in this work, we somewhat doubt their findings of a stable 2-planet system and question whether such a system is really supported by the new data set given that no strict confidence levels were found for the best-fit outer period. Since performing a full re-analysis of their newly compiled data, including dynamical mapping of their new architecture, would be a particularly time intensive process, we have chosen to postpone this task for a future study.

\vspace{-0.3cm}
\section*{Acknowledgments}

\begin{figure*}
\includegraphics[scale=1.0]{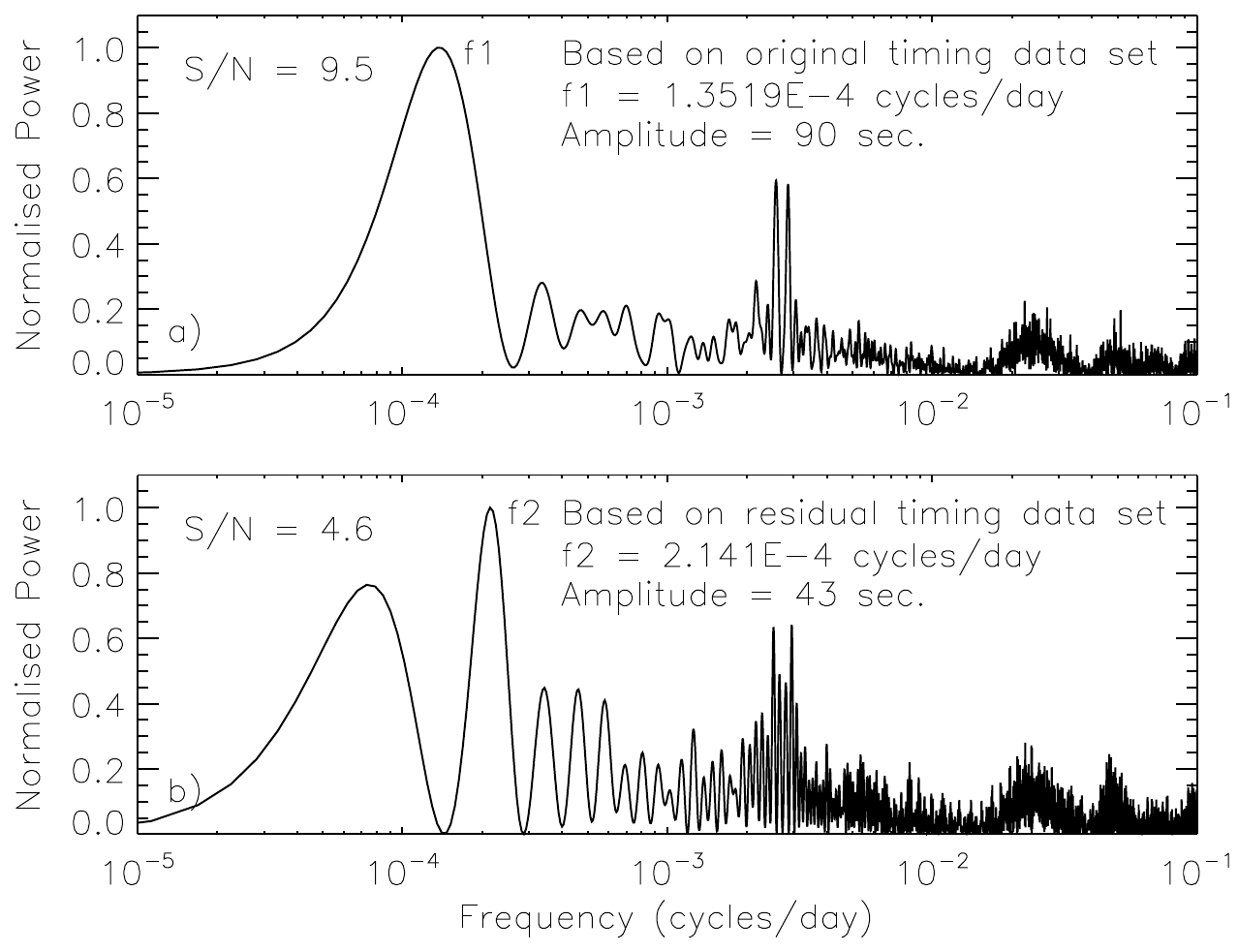}
\caption{Power spectrum of HW~Vir timing data set using BJD(TDB) times with the $O-C$ residuals measured in seconds. Two periods were found with 
$1/f_1 = 7397$ and $1/f_2 = 4672$ days, corresponding to 20.3 and 
12.8~years, respectively.  S/N denotes the frequency signal-to-noise 
ratio, which are significantly larger than the spectrum's noise level.  
Normalisation was done by division of the maximum amplitude in each 
spectrum. The $f_2$-frequency was determined from the residuals after 
subtracting the period $1/f_1$ from the original timing data set.  
Additional peaks in both panels represent 1-year alias frequencies due 
to the repeating annual observing cycle of HW~Vir.}
\label{HWVirPowerSpec}
\end{figure*}

\begin{figure*}
\includegraphics[scale=0.55]{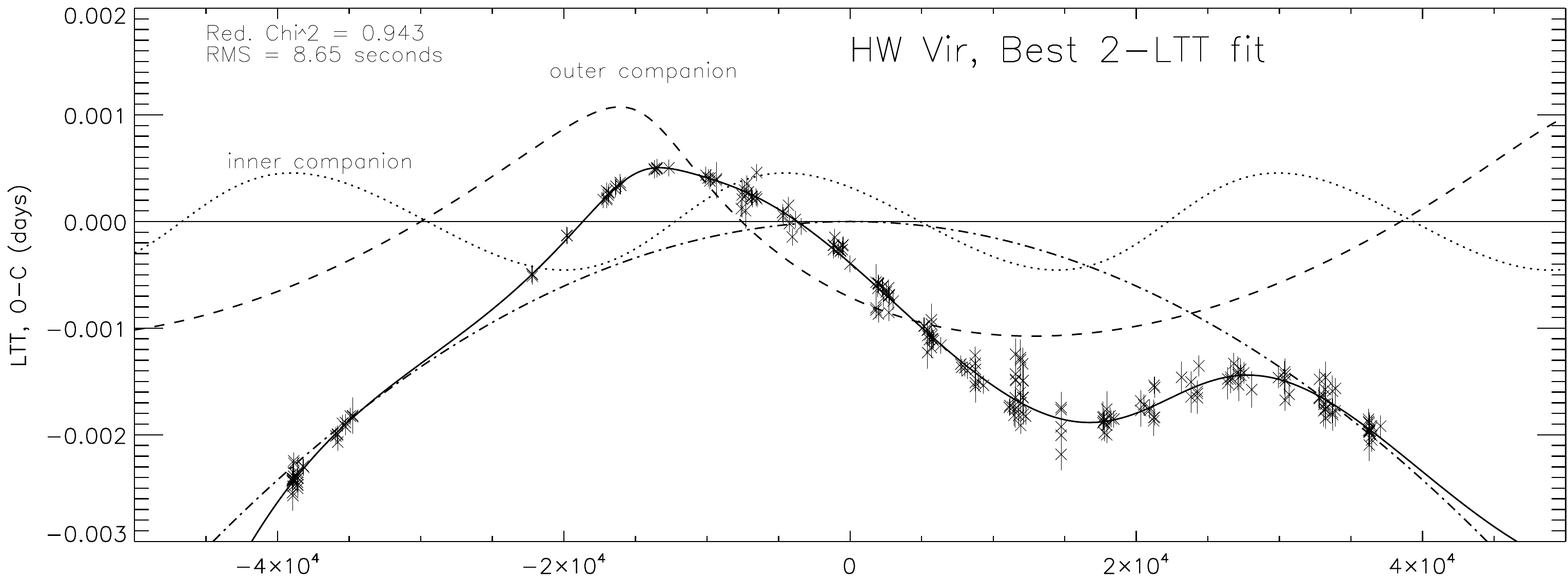}
\includegraphics[scale=0.55]{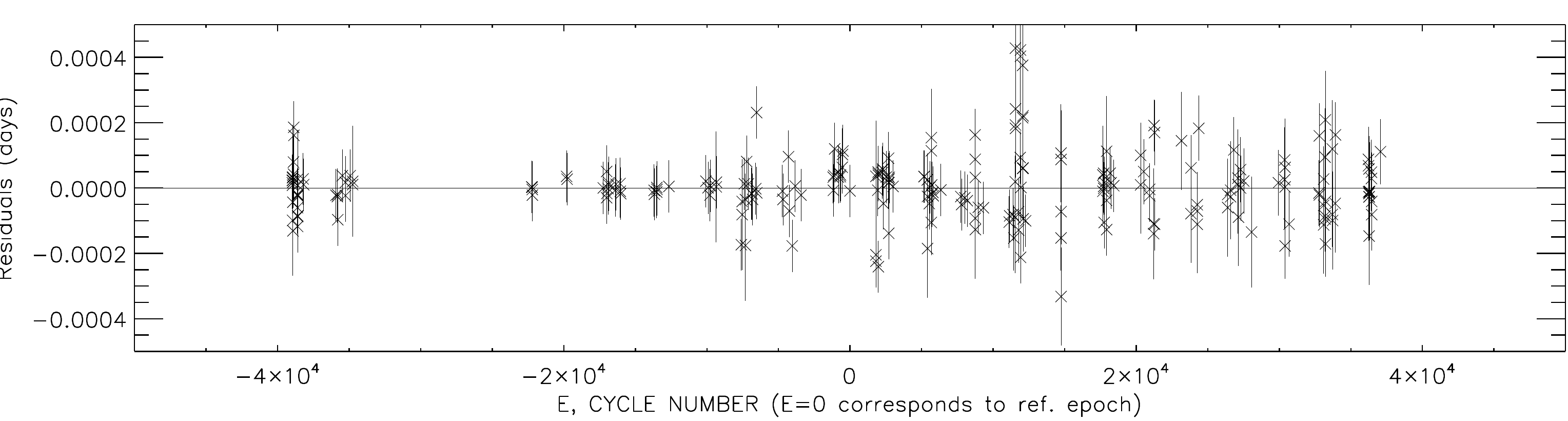}
\caption{Best-fit two-Kepler LTT model with $\chi_{r}^2=0.943$.  
The best-fit model parameters (including reference epoch) are shown in Table~\ref{bestfitparamtable}.RMS denotes the root-mean-square scatter around the best fit. The lower part of the figure shows the residuals between the best fit model and the observed timing data set.}
\label{HWVirBestFit}
\end{figure*}

\begin{figure*}
\vbox{
  \hbox{\includegraphics[width=0.25\textwidth, angle=270]{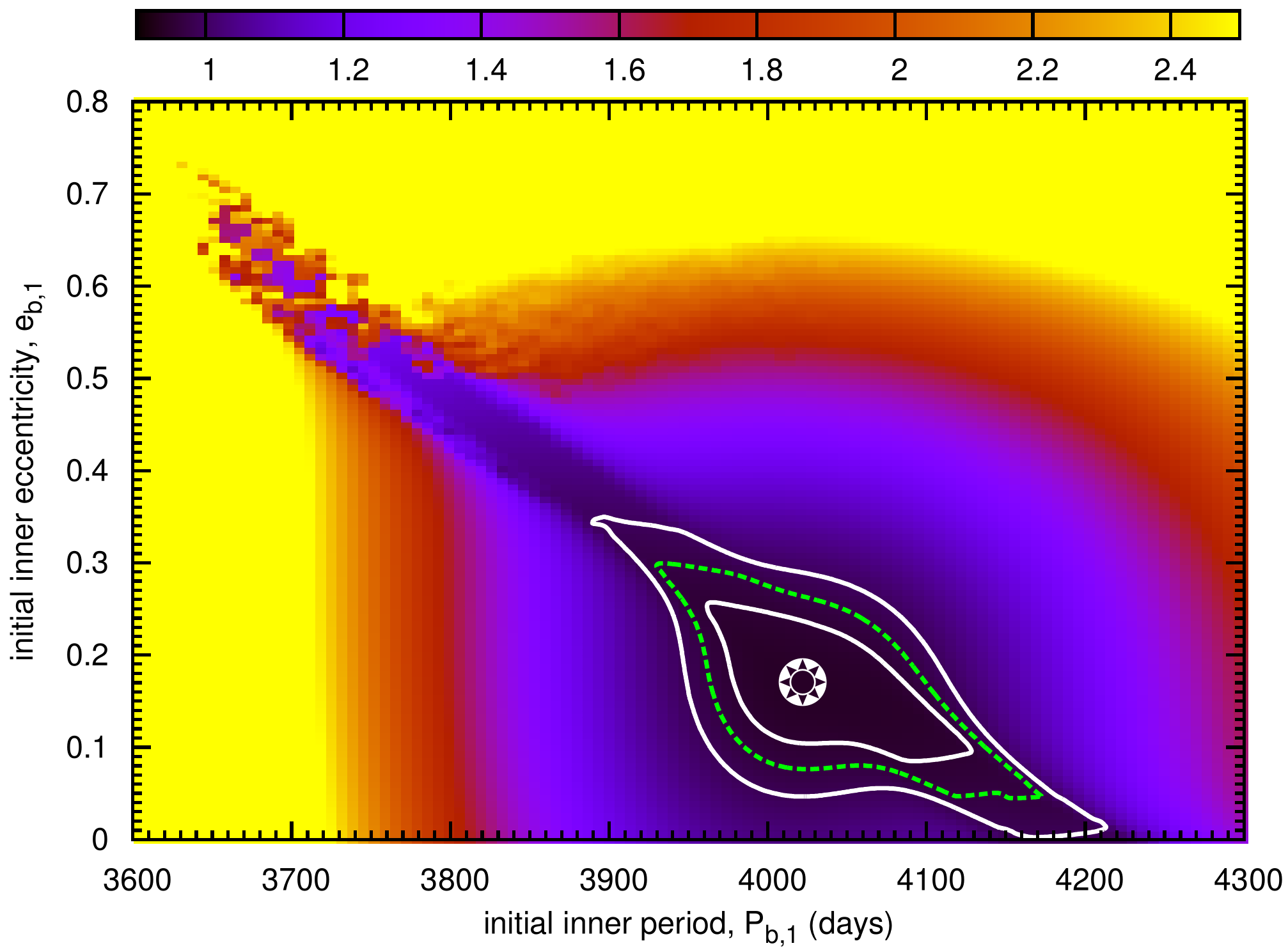}
        \includegraphics[width=0.25\textwidth, angle=270]{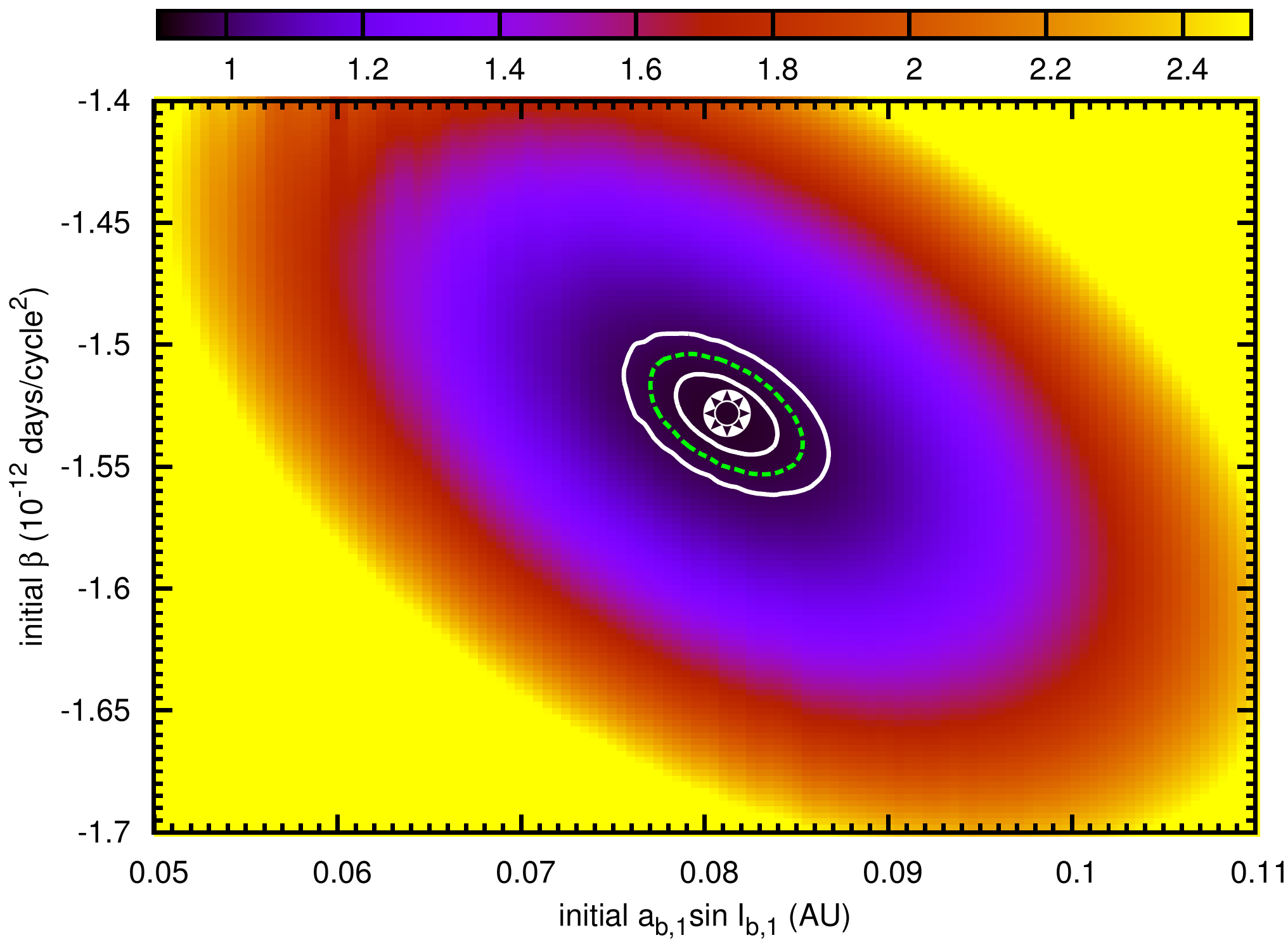}
        \includegraphics[width=0.25\textwidth, angle=270]{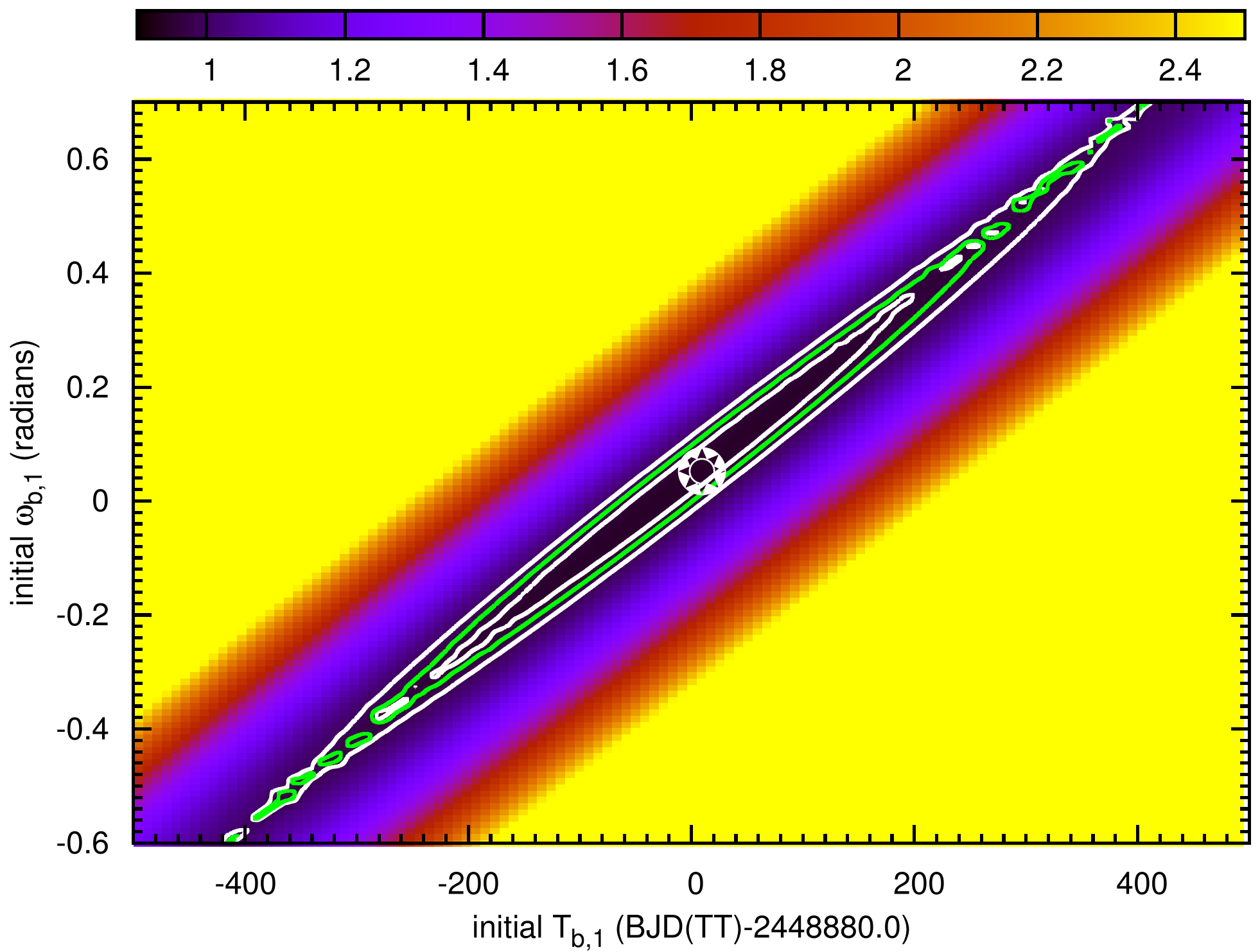}
  }
  \hbox{\includegraphics[width=0.245\textwidth, angle=270]{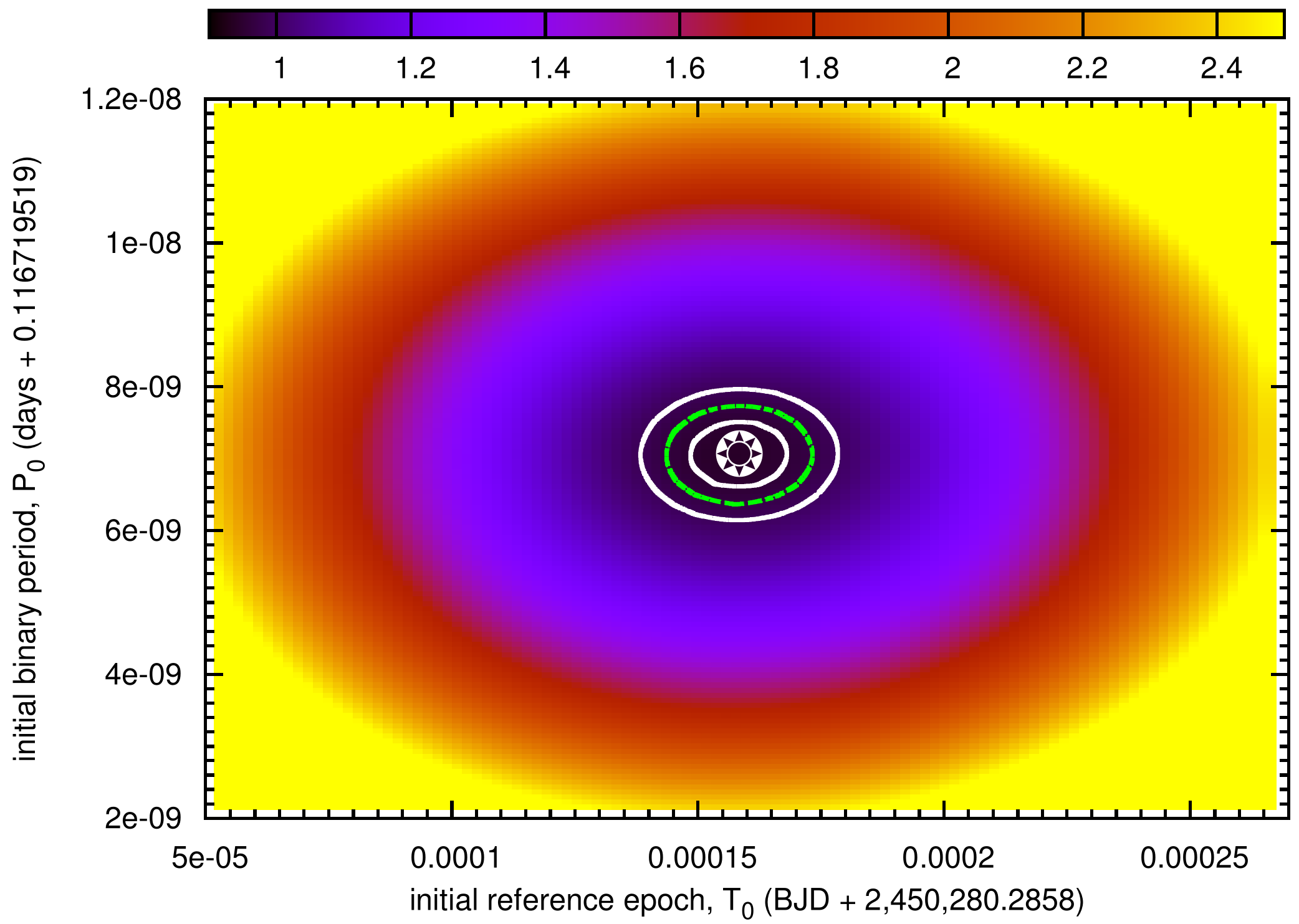}
        \includegraphics[width=0.245\textwidth, angle=270]{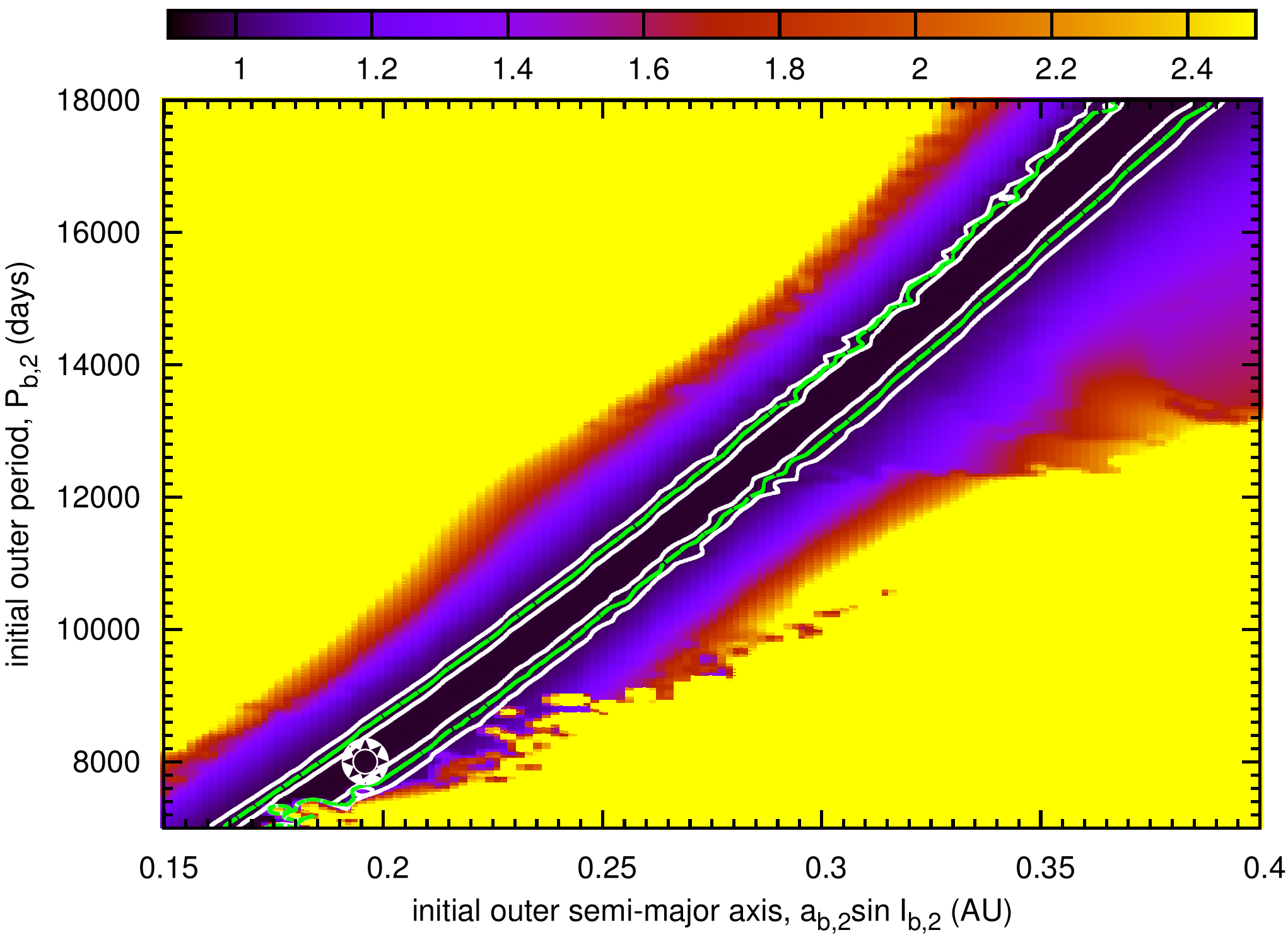}
        \includegraphics[width=0.245\textwidth, angle=270]{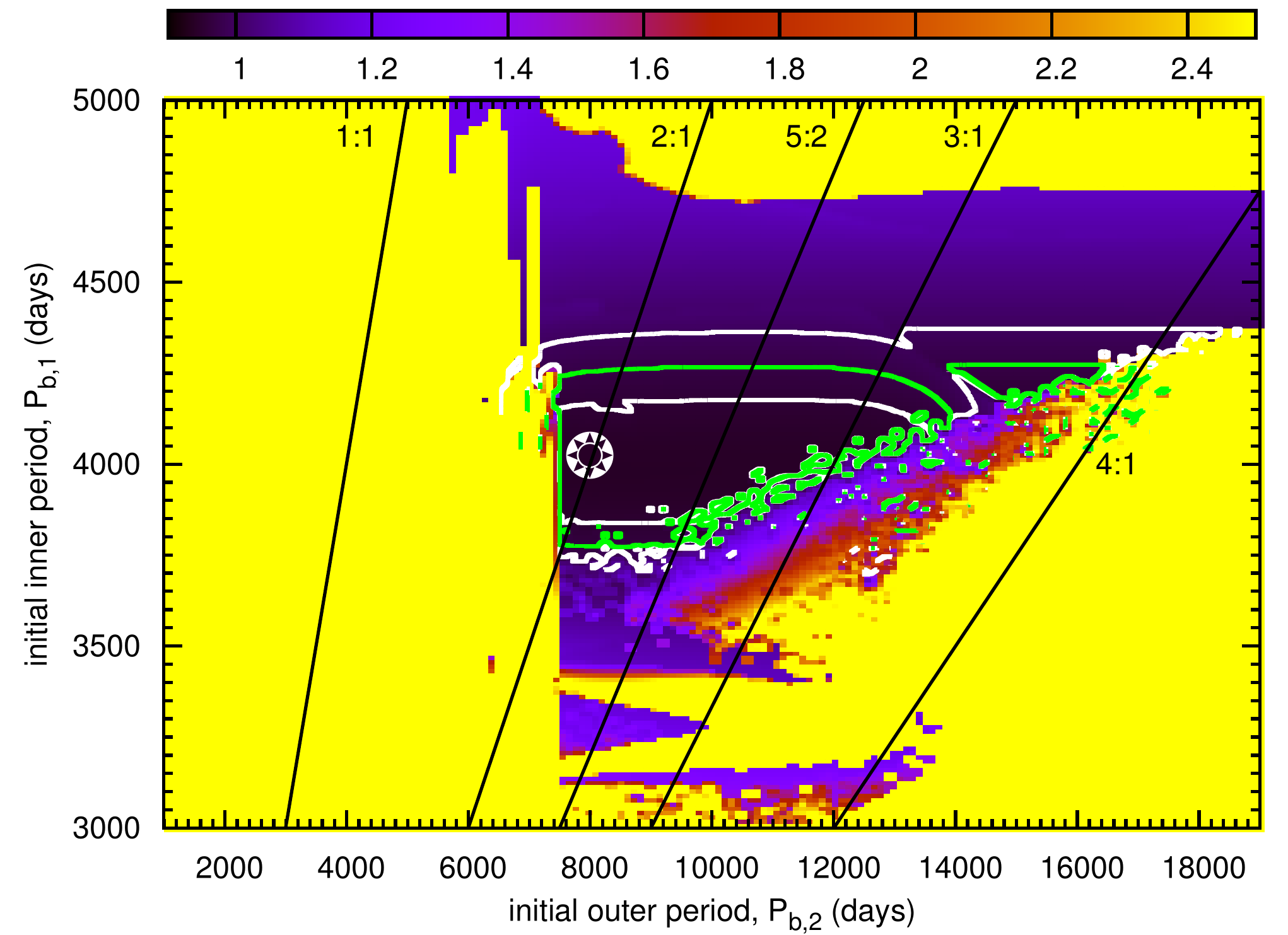}

  }
}
\vspace{0.0cm}
\caption{Colour-coded $\chi_{r}^2$ parameter scans of orbital parameters with remaining 
parameters to vary freely. The best-fit parameter is indicated by a star-like symbol. Contour curves show the 1,2,3$\sigma$ confidence level curves around our best-fit model. 
See electronic version for colour figures.}
\label{3by2plot}
\end{figure*}

\begin{figure*}
\vbox{
  \hbox{\includegraphics[width=0.25\textwidth, angle=270]{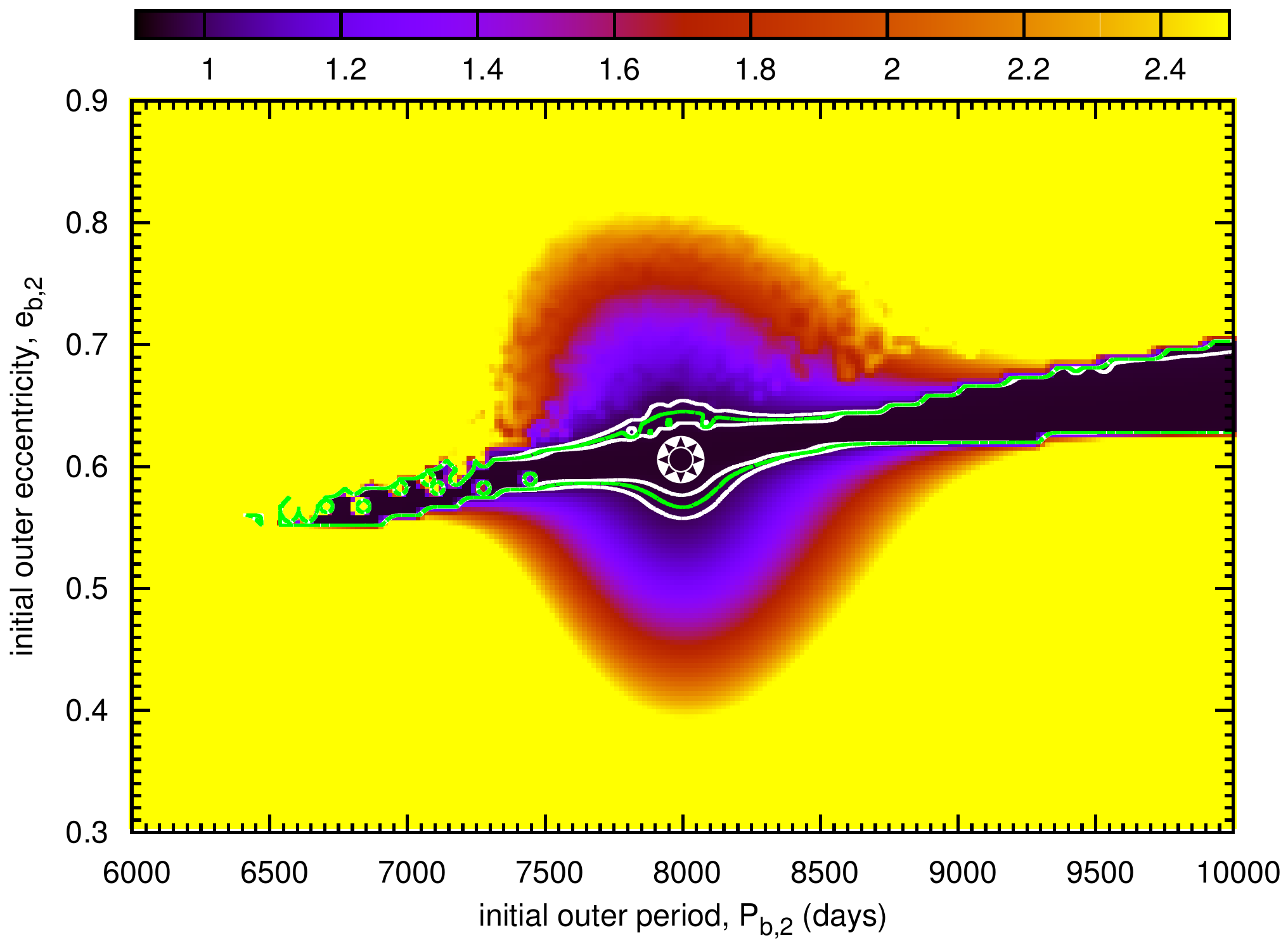}
        \includegraphics[width=0.25\textwidth, angle=270]{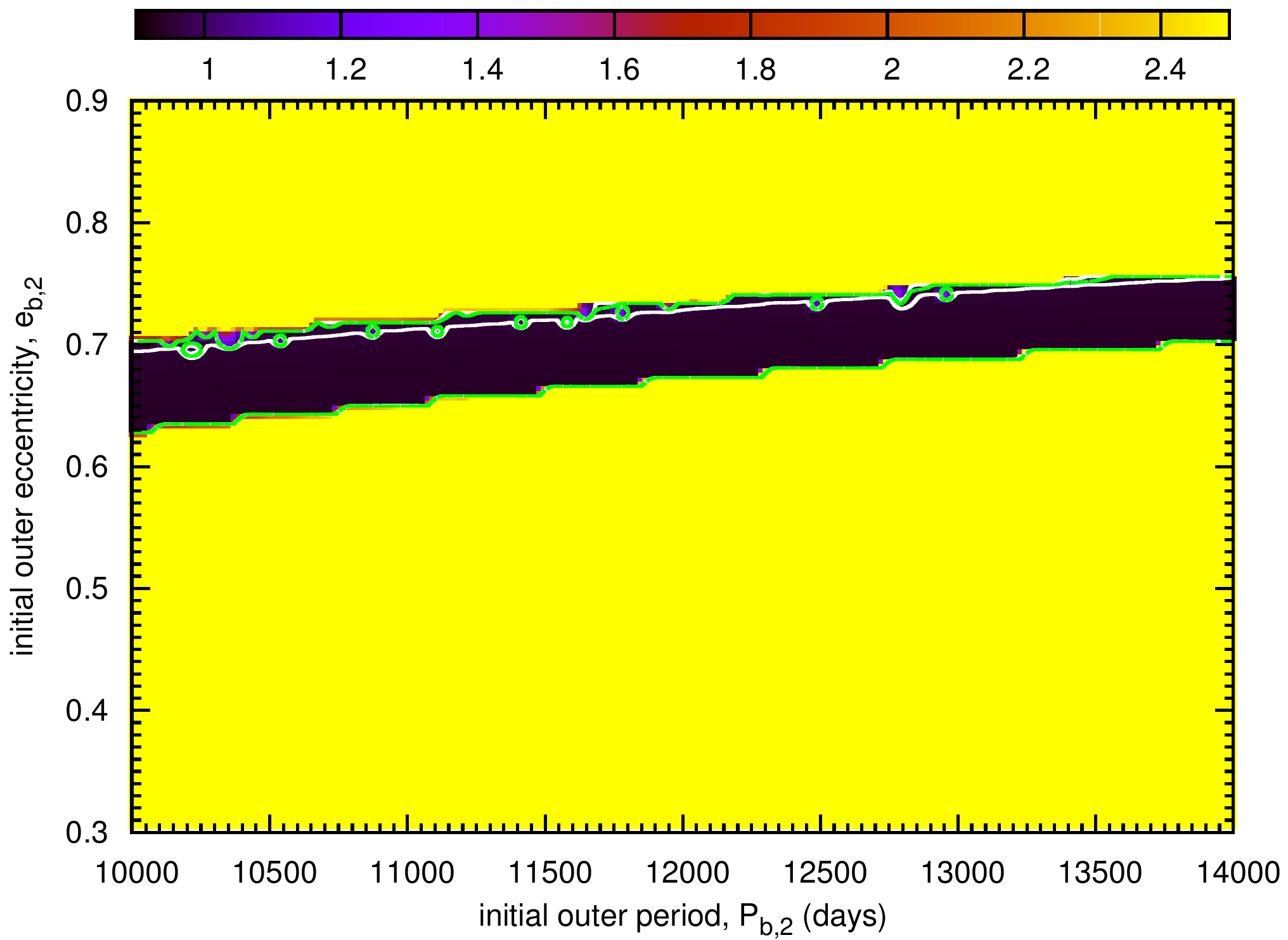}
        \includegraphics[width=0.25\textwidth, angle=270]{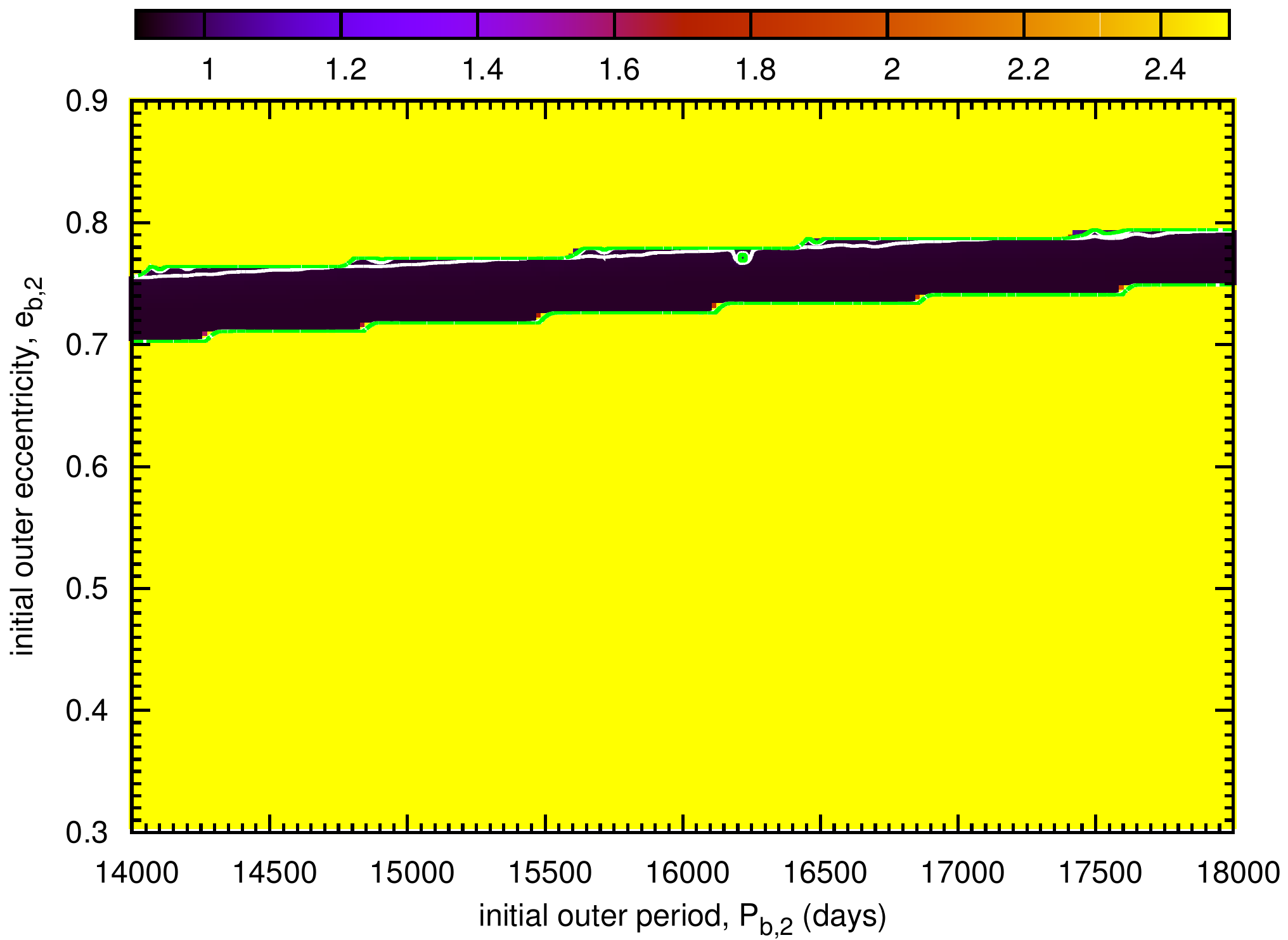}
  }
}
\vspace{0.0cm}
\caption{Same as Fig.~\ref{3by2plot} but considering the $(P_{2},e_{2})$-plane. The middle and right panel shows the $\chi^2_r$ topology for longer orbital periods of the outer companion. Based on the used data set, no firm confidence levels can be established around out best-fit value.}
\label{3by1plot}
\end{figure*}

\begin{figure*}
\vbox{
  \hbox{\includegraphics[width=0.335\textwidth, angle=0]{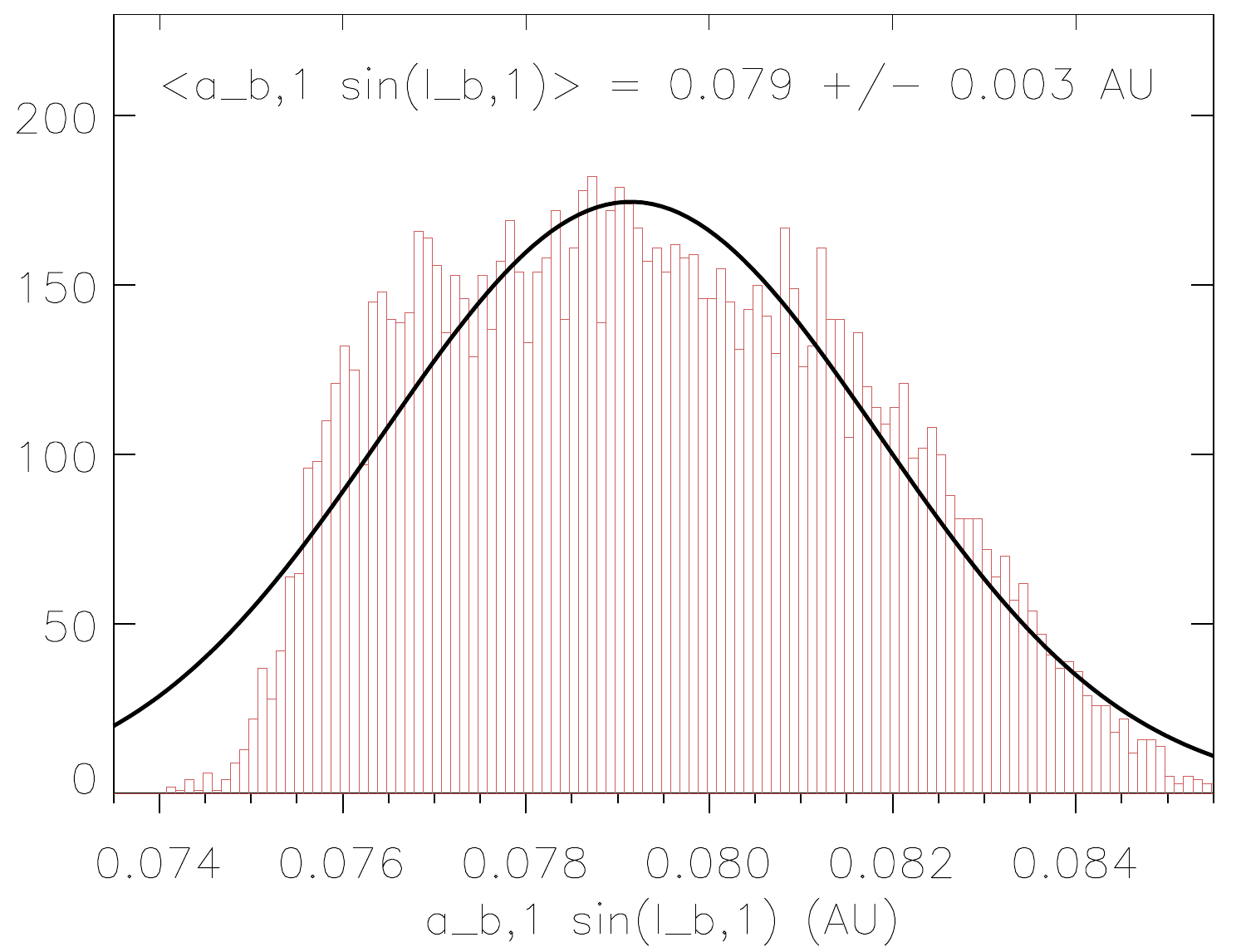}
        \includegraphics[width=0.335\textwidth, angle=0]{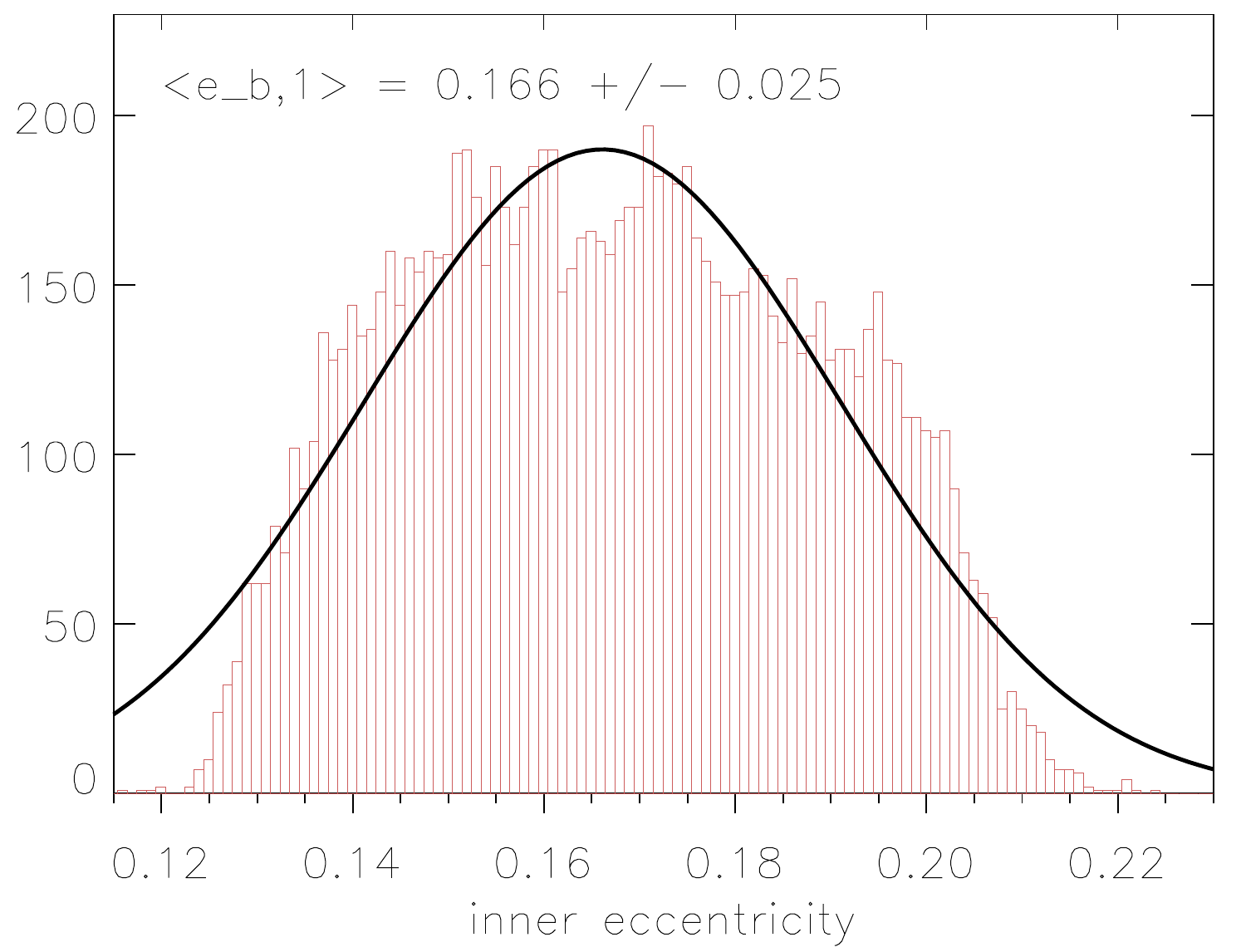}
        \includegraphics[width=0.335\textwidth, angle=0]{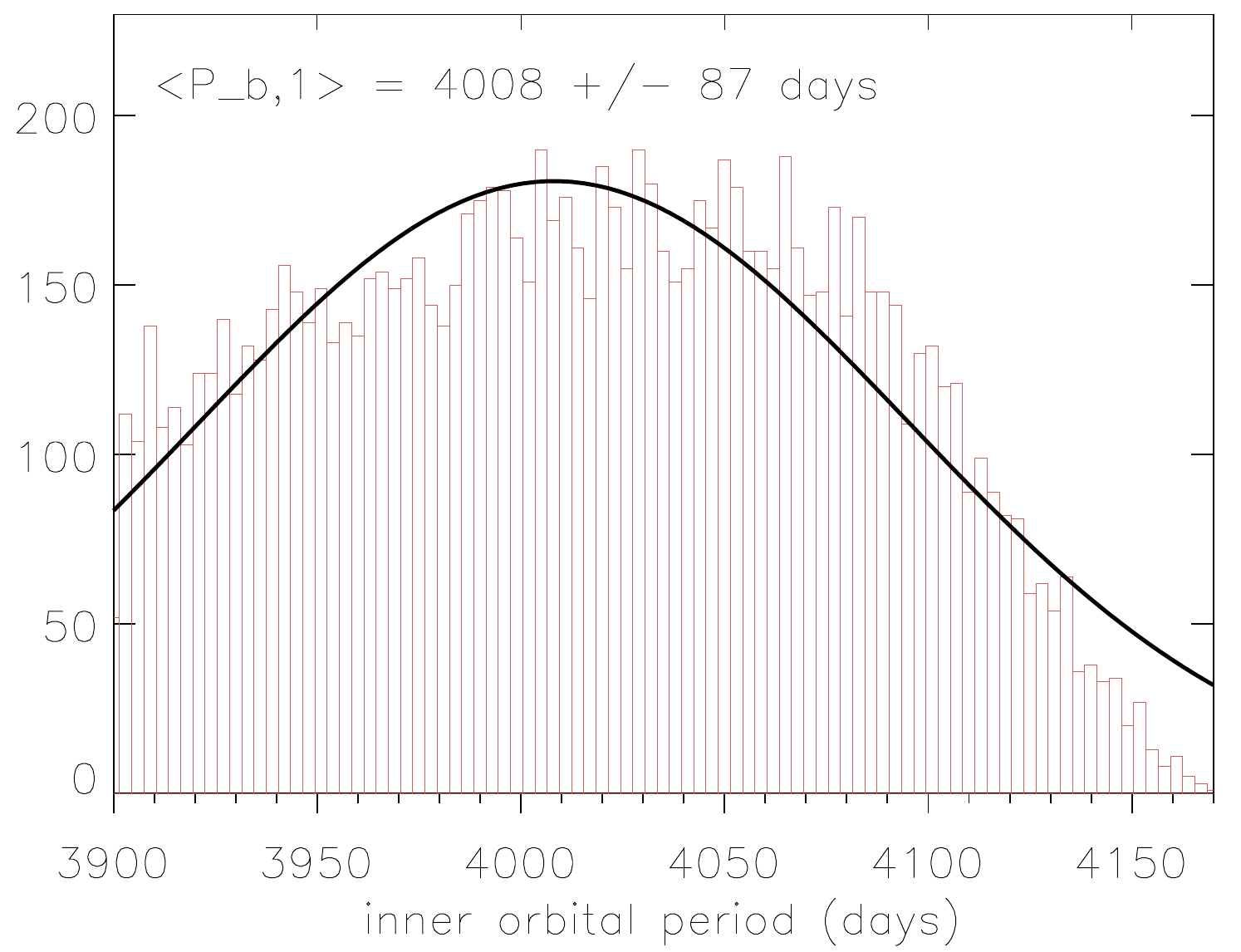}
  }
  \hbox{\includegraphics[width=0.335\textwidth, angle=0]{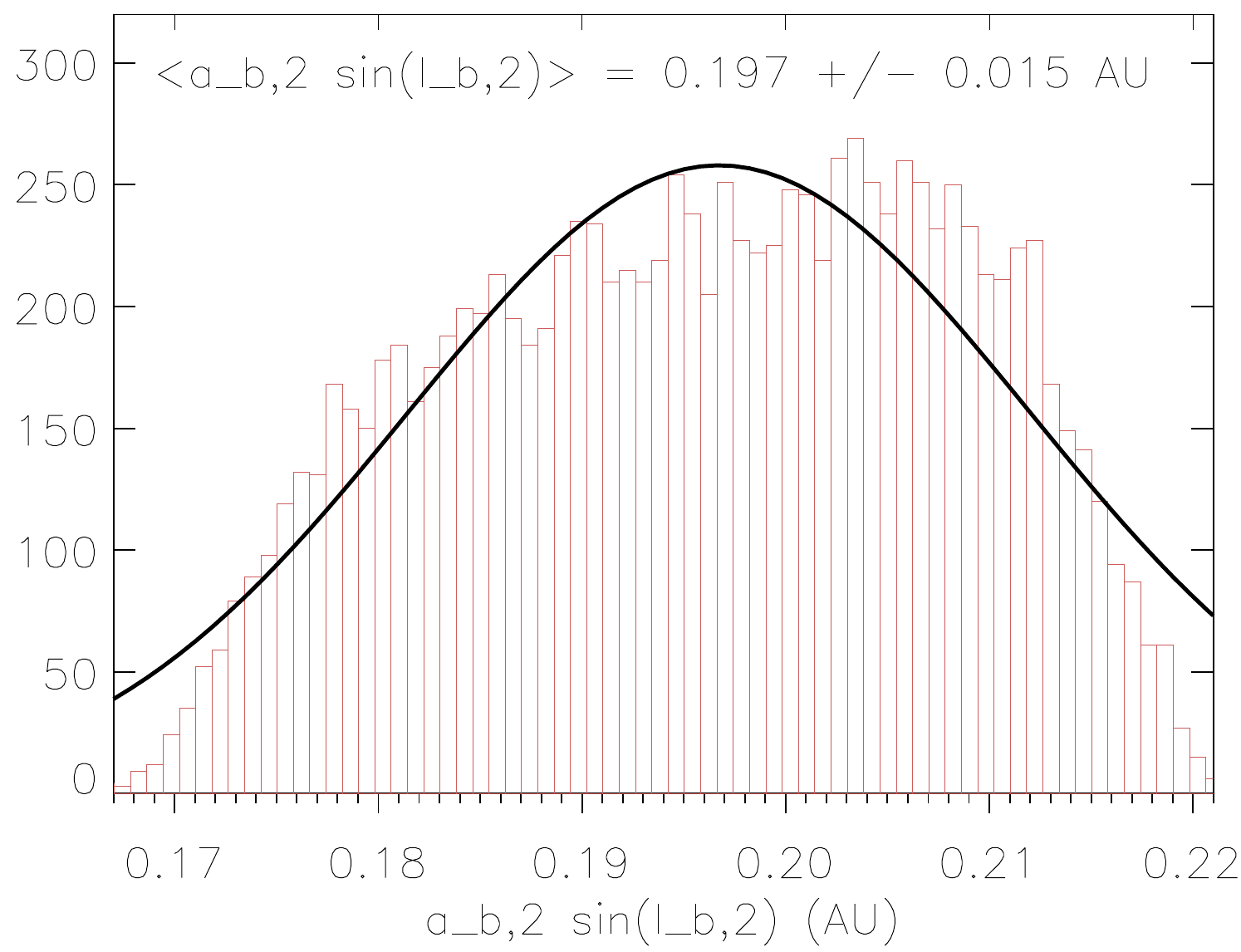}
        \includegraphics[width=0.335\textwidth, angle=0]{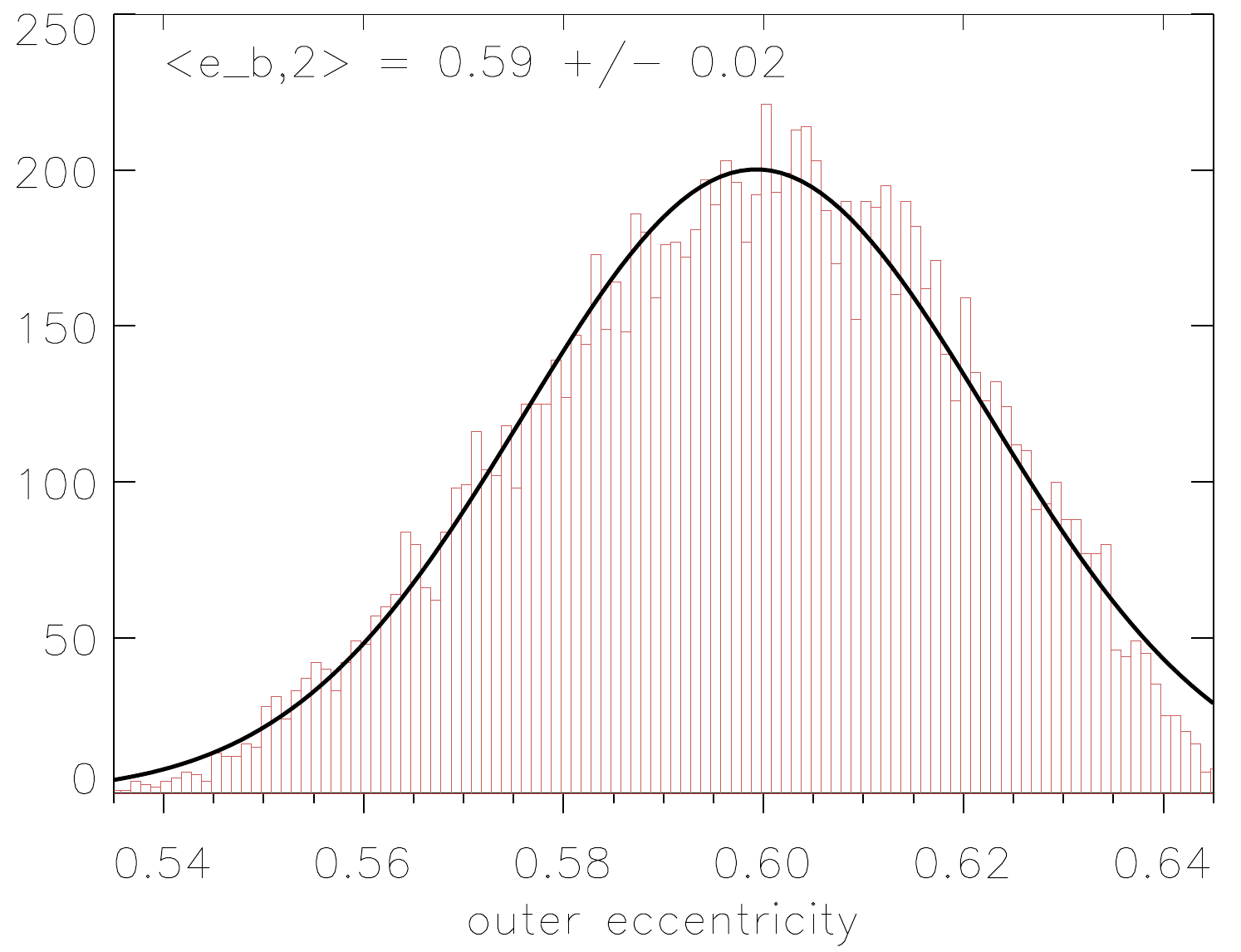}
        \includegraphics[width=0.339\textwidth, angle=0]{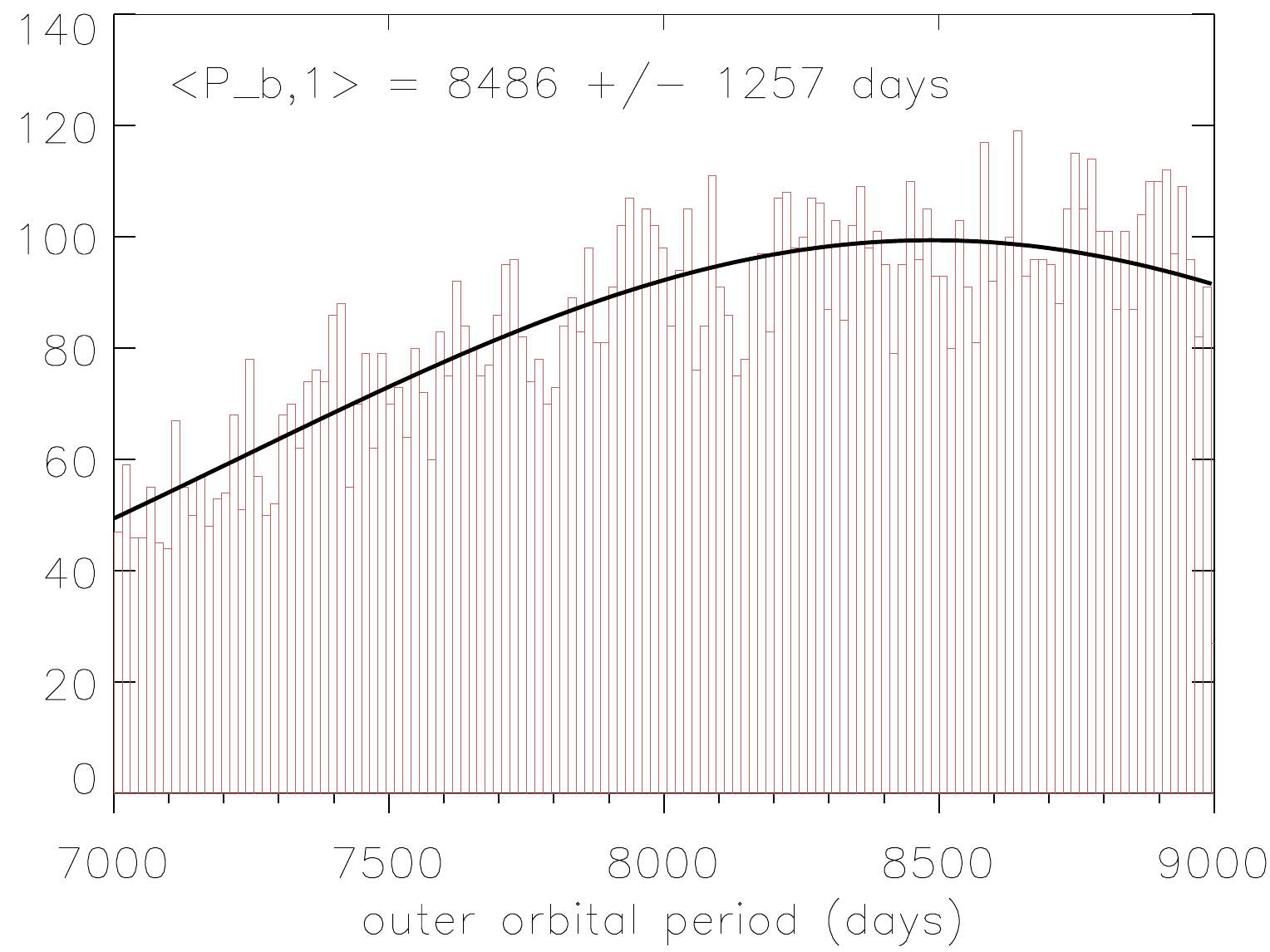}
  }
}
\vspace{0.0cm}
\caption{Histogram distribution of six model parameters as obtained from a Monte Carlo 
experiment. Only models with $\chi^{2}_{r} < 1-\sigma$ where considered to assess the 68\% confidence levels for each parameter. Solid curves show fitted normal distribution with mean and standard deviation indicated in each panel. However, we used the mean and standard deviation derived from the underlying dataset to derive our 1-$\sigma$ errors as quoted in Table~\ref{bestfitparamtable}.}
\label{ErrorHistogram}
\end{figure*}

\begin{figure*}
\includegraphics[scale=0.8]{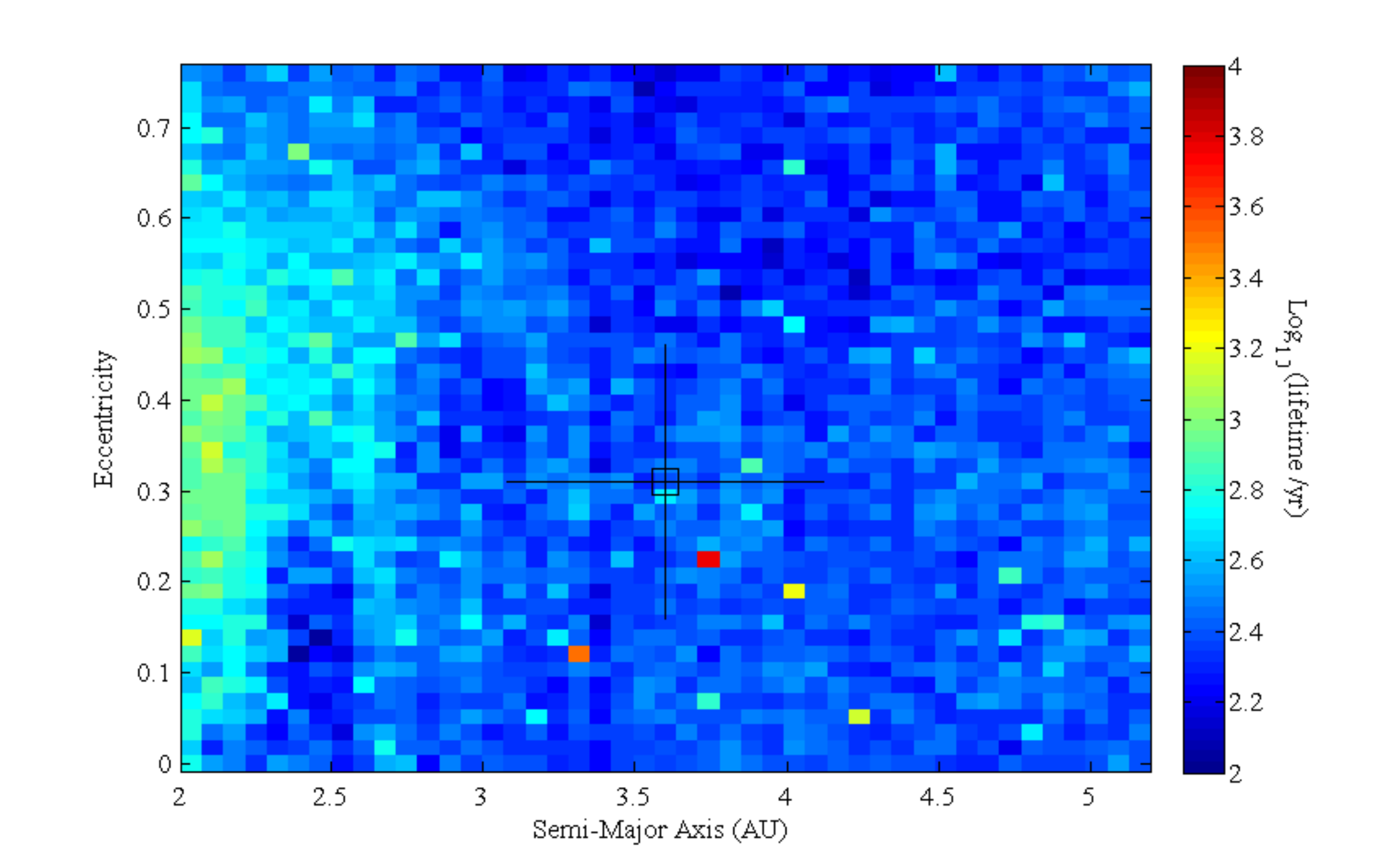}
\caption{The stability of the HW~Vir planetary system as proposed by 
\citet{Lee2009}, as a function of the semi-major axis, $a$, and 
eccentricity, $e$, of planet HW~Vir~b.  The initial orbit of 
HW~Vir~c was the same in each integration, set to the nominal best 
fit orbit from that work.  The mean lifetime of the planetary system (in 
$log_{10} (lifetime /yr)$) at a given $a-e$ co-ordinate is denoted by the 
colour of the plot.  The lifetime at each $a-e$ location is the mean 
value of 45 separate integrations carried of orbits at that $a-e$ 
position (testing a combination of 15 unique $\omega$ values, and 3 
unique $M$ values).  The nominal best-fit orbit for HW~Vir~b is 
located within the open square, from which lines radiate showing the 
extend of the $\pm~1$-$\sigma$ errors on $a$ and $e$.  As can be seen, the 
orbits of the system are incredibly unstable, no matter what initial 
orbit is considered for HW~Vir b.}
\label{OriginalStability}
\end{figure*}

\begin{figure*}
\includegraphics[scale=0.7]{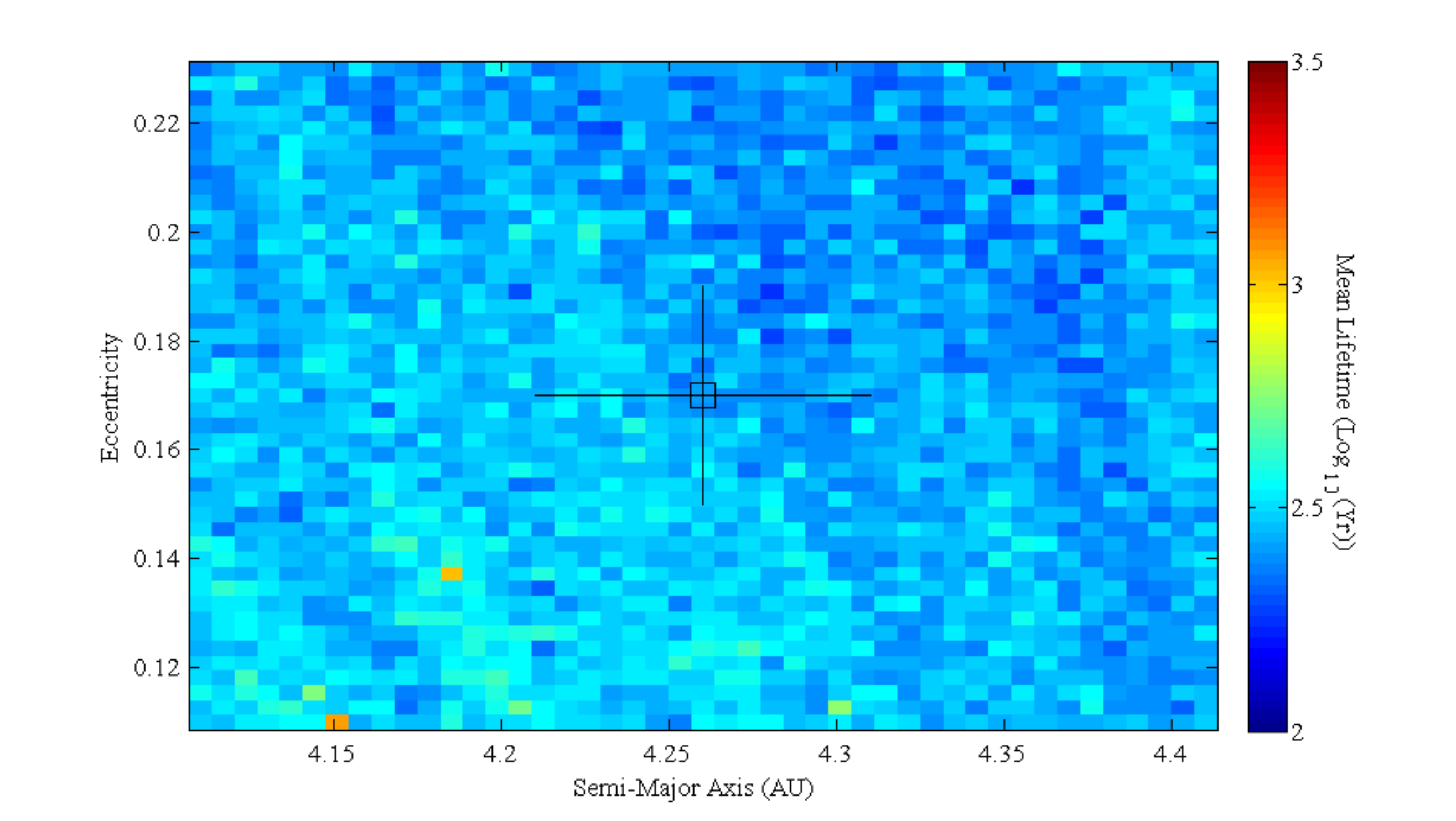}
\caption{The stability of the HW~Vir planetary system, given the orbital 
solution derived in this work, as a function of the semi-major axis, 
$a$, and eccentricity, $e$, of planet HW~Vir~b.  The initial orbit of 
HW~Vir c was the same in each integration, set to the nominal best 
fit orbit as detailed in Table 2.  The mean lifetime of the planetary system 
(shown as $log_{10} (lifetime /yr$) at a given $a-e$ co-ordinate is denoted by 
the colour of the plot.  The lifetime at each $a-e$ location is the mean 
value of 45 separate integrations carried of orbits at that $a-e$ 
position (testing a combination of 15 unique $\omega$ values, and 3 
unique $M$ values).  The nominal best-fit orbit for HW~Vir~b is 
located within the open square, from which lines radiate showing the 
extend of the $\pm~1$-$\sigma$ errors on $a$ and $e$.  Once again, the 
orbits of the system are found to be incredibly unstable, no matter what 
initial orbit is considered for HW~Vir~b. The
two red hotspots in that plot are the result of two unusually stable
runs, with lifetimes of 33~kyr ($a = 4.185$ AU, $e = 0.137$) and 38~kyr
($a = 4.15$ AU, $e = 0.11$). Even these most extreme outliers are
dynamically unstable on astronomically short timescales.}
\label{DynamicsNew}
\end{figure*}

The authors wish to thank an anonymous referee, whose extensive comments on our work
led to significant changes that greatly improved the depth of this work. 
We also thank Dr.~Lee Jae Woo for useful discussions and suggestions that 
resulted in the creation of Fig.~\ref{HWVirPowerSpec}. 
JH gratefully acknowledges the 
financial support of the Australian government through ARC Grant 
DP0774000.  RW is supported by a UNSW Vice-Chancellor's Fellowship.  JPM 
is partly supported by Spanish grants AYA 2008/01727 and AYA 2011/02622..  TCH gratefully 
acknowledges financial support from the Korea Research Council for 
Fundamental Science and Technology (KRCF) through the Young Research 
Scientist Fellowship Program, and also the support of the KASI (Korea 
Astronomy and Space Science Institute) grant 2012-1-410-02.  The 
dynamical simulations performed in this work were performed on the EPIC 
supercomputer, supported by iVEC, located at the Murdoch University, in 
Western Australia. The Monte Carlo/fitting simulations were carried out on the 
``Beehive'' computing cluster at Armagh Observatory (UK) and the ``Pluto'' high performance computing cluster at the Korea Astronomy and Space Science 
Institute. Astronomical research at Armagh Observatory (UK) is funded by the Department of Culture, Arts and Leisure.

\vspace{-0.6cm}
\bibliographystyle{mn_new}

\clearpage

\clearpage

\clearpage 

\clearpage

\begin{table*}
\centering
\begin{tabular}{lcccc}
\hline
Parameter && \multicolumn{2}{c}{two-LTT} & Unit \\ [1.5mm] \cline{3-4} \\ [-2.0ex]
                         && $\tau_{1}~(i=1)$                 & $\tau_{2}~(i=2)$     &   \\ 
\hline
$\chi_{r}^2$            &&  \multicolumn{2}{c}{0.943}                          &   -   \\
\hline
RMS            &&  \multicolumn{2}{c}{8.665}                                     & seconds   \\
\hline
$\beta$ && \multicolumn{2}{c}{$-1.529\cdot 10^{-12} \pm 1.25\cdot 10^{-13}$}   & $\textnormal{day/cycle}^{2}$   \\
\hline
$T_0$ && \multicolumn{2}{c}{$2,450,280.28596\pm 2.3\cdot 10^{-5}$}             & BJD(TDB) \\
$P_0$ && \multicolumn{2}{c}{$0.116719519\pm 4.6 \cdot 10^{-9}$}                  & days \\
$a_{b,i}\sin I_{b,i}$                && $0.081 \pm 0.002$ & $0.196 \pm 0.012$ & AU   \\
$e_{b,i}~~(\textnormal{or}~e_{1,2})$ && $0.17 \pm 0.02$   & $0.61 \pm 0.02$      & -    \\
$\omega_{b,i}$                       && $0.05 \pm 0.01$       & $2.09 \pm 0.08$   & rad.  \\
$T_{b,i}$                            && $2,448,880 \pm 57$  & $2,448,629 \pm 42$     & BJD(TT) \\
$P_{b,i}~~(\textnormal{or}~P_{1,2})$ && $4021 \pm 64$    & $7992 \pm 551$ (!)       & days \\
\hline
$K_{b,i}$&&$4.6\cdot 10^{-4}\pm 1.3\cdot 10^{-5}$&$1.13\cdot 10^{-3}\pm 7.04\cdot 10^{-5}$&days\\
\hline
$m_{i} \sin I_{i}$                   && $12 \pm 3$            & $11 \pm 8 $ & $M_{Jup}$ \\
$a_{i}\sin I_{i}$                    && $4.26 \pm 0.05$            & $6.8 \pm 0.3$ & AU   \\
$e_{i}$                              && $0.17 \pm 0.02$   & $0.61 \pm 0.02$      & -    \\
$\omega_{i}$                         && $(\pi - 0.05) \pm 0.01$       & $(\pi - 2.09) \pm 0.08$           & rad.  \\
$T_{i}$  && $2,448,880 \pm 57$  & $2,448,629 \pm 42$     & BJD(TT) \\
$P_{i}$                              && $4021 \pm 64$    & $7992 \pm 551$ (!)       & days \\
\hline
\end{tabular}
\caption{Best-fit parameters for the LTT orbits of HW~Vir corresponding 
to Fig.~\ref{HWVirBestFit}. Subscripts $1,2$ refer to the circumbinary 
companions with $i=1$, the inner, and $i=2$, the outer, companions.  RMS 
measures the root-mean-square scatter of the data around the best fit.  
1-$\sigma$ uncertainties have been obtained as described in the text. The last 
five entries are quantities of the two companions in the astrocentric 
coordinate system. Note that our values for $a$ and $P$ are somewhat 
larger than those of \citet{Lee2009}: $a_b=3.62$ and $a_c=5.30$ AU, 
$P_b=3316$ and $P_c=5786$ days.}
\label{bestfitparamtable}
\end{table*}

\label{lastpage}	
\end{document}